\documentclass[12pt]{article}
\usepackage[]{color}
\pdfoutput=1

\usepackage{amsmath}
\usepackage{amssymb}
\usepackage{amsthm}
\usepackage[english]{babel}
\usepackage{graphicx}
\usepackage{hyperref}  
\usepackage[authoryear,comma,longnamesfirst,sectionbib]{natbib} 
\usepackage{mathtools}
\usepackage{threeparttable}
\usepackage{xcolor}
\usepackage{tikz}
\usepackage{caption}
\usepackage{booktabs}
\usepackage{enumitem}
\usepackage[affil-it]{authblk}
\usepackage{lscape}
\usepackage[margin=1in]{geometry}
\mathtoolsset{showonlyrefs=true} 
\DeclareMathOperator{\vc}{vec}

\usepackage{setspace}

\newif\iffig
\figtrue
\title{A Flexible Mixed-Frequency Vector Autoregression with a Steady-State Prior\footnote{Corresponding author: Sebastian Ankargren. E-mail: \href{mailto:sebastian.ankargren@statistics.uu.se}{sebastian.ankargren@statistics.uu.se}}}
\author[1]{Sebastian Ankargren}
\author[1]{M\r{a}ns Unosson}
\author[1,2]{Yukai Yang}
\affil[1]{Department of Statistics, Uppsala University}
\affil[2]{HHS}
\date{}                     

\usepackage{tabularx}

\usepackage[symbol]{footmisc}

\begin{document}

\newcommand{\under}[1]{\underbar{$ #1 $}}

\begin{titlepage}
\pagestyle{empty}
\vspace*{\fill}
  \vskip 2em%
  \begin{center}%
  \let \footnote \thanks
    {\LARGE 
\begin{spacing}{1.2}A Flexible Mixed-Frequency Vector Autoregression with a Steady-State Prior\textsuperscript{*}
\end{spacing}}%
    \vskip 1.5em%
    {\large
      \lineskip .5em%
      \begin{tabular}[t]{c}%
        Sebastian Ankargren$^{1,\dagger}$, M\r{a}ns Unosson$^1$, and Yukai Yang$^{1,2}$\vspace{0.25cm}\\
        $^1$\emph{Department of Statistics, Uppsala University}\\
        $^2$\emph{Center for Data Analytics, Stockholm School of Economics}\\
      \end{tabular}\par}%
    \vskip 1em%
  \end{center}%
  \par
  \vskip 1.5em

\begin{abstract}
We propose a Bayesian vector autoregressive (VAR) model for mixed-frequency data. Our model is based on the mean-adjusted parametrization of the VAR and allows for an explicit prior on the `steady states' (unconditional means) of the included variables. Based on recent developments in the literature, we discuss extensions of the model that improve the flexibility of the modeling approach. These extensions include a hierarchical shrinkage prior for the steady-state parameters, and the use of stochastic volatility to model heteroskedasticity. We put the proposed model to use in a forecast evaluation using US data consisting of 10 monthly and 3 quarterly variables. The results show that the predictive ability typically benefits from using mixed-frequency data, and that improvements can be obtained for both monthly and quarterly variables. We also find that the steady-state prior generally enhances the accuracy of the forecasts, and that accounting for heteroskedasticity by means of stochastic volatility usually provides additional improvements, although not for all variables.
\end{abstract}

\footnotetext[1]{This work has been supported by Jan Wallanders och Tom Hedelius stiftelse samt Tore Browaldhs stiftelse, grant number P2016-0293:1. The computations were performed on resources provided by SNIC through Uppsala Multidisciplinary Center for Advanced Computational Science (UPPMAX) under Project SNIC 2015/6-117.}
\footnotetext[2]{Corresponding author: \mbox{S. Ankargren,~\href{mailto:sebastian.ankargren@statistics.uu.se}{sebastian.ankargren@statistics.uu.se}}}
\vspace*{\fill}

\end{titlepage}

\section{Introduction}
The vector autoregressive model (VAR) is a commonly used tool in applied macroeconometrics, in part because of its simplicity. Over the years, VAR models have developed in many different directions under both frequentist and Bayesian paradigms. The Bayesian approach offers the attractive ability to easily incorporate soft restrictions and shrinkage, which ameliorate the issue of overparametrization. Within the Bayesian framework itself, a large number of papers have developed prior distributions for the parameters in VAR models. Many of these are, in one way or another, variations of the Minnesota prior proposed by \cite{Litterman1986} (see for example the book chapters \citealp{DelNegro2011,Karlsson2013}). Gains in computational power have led to further alternatives in the choice of prior distribution as intractable posteriors can efficiently be sampled using Markov Chain Monte Carlo (MCMC) methods such as the Gibbs sampler \citep{Gelfand1990,Kadiyala1997}.

A particular development in the Bayesian VAR literature is the steady-state prior proposed by \citet{Villani2009}. The prior is based on a mean-adjusted form of the VAR where the unconditional mean is explicitly parameterized. This seemingly innocuous reparametrization is justified by the fact that practitioners and analysts often have prior information regarding the steady-state (or unconditional mean) readily available. In the standard parametrization, a prior on the unconditional mean is only implicit as a function of the other parameters' priors. Because the forecast in a stationary VAR converges to the unconditional mean as the horizon increases, a prior for the steady-state parameters can help retain the long-run forecast in the direction implied by theory, even if the model is estimated during a period of divergence.

Another modeling feature that modern VARs often include is stochastic volatility. In many macroeconomic applications, a typical characteristic of the data is that the volatility has varied over time. By fitting VARs with constant volatility, the estimated error covariance matrix attempts to balance periods of low and high volatility and find a compromise. Consequently, the predictive distribution does not account for the current level of volatility. Seminal contributions with respect to stochastic volatility were first made by \cite{Primiceri2005,Cogley2005} and numerous follow-up studies have since documented the usefulness of stochastic volatility for forecasting, see e.g. work by \cite{Clark2011,DAgostino2013,Clark2013,Carriero2016}. Because of the established utility thereof, we also allow for more flexibility in our model by modeling time variation in the error covariance matrix.

VARs are often estimated on a quarterly basis, see e.g. \cite{Stock2001,Adolfson2007b}. The reason is simply that many variables of interest are unavailable at higher frequencies, although the majority is often sampled monthly if not even more frequently. When the data are available at different frequencies, common practice is to aggregate high-frequency variables to the lowest frequency present. Such an aggregation incurs a loss of information for variables measured throughout the quarter: the aggregated quarterly values are typically weighted sums of the constituent months and so any information carried by a within-quarter trend or pattern will be disregarded by the aggregation.
From a forecasting perspective an analyst will be unconsciously forced to disregard part of the information set when constructing a forecast from within a quarter as the most recent realizations are only available for the high-frequency variables. Another reason for utilizing higher frequencies of the data is that the number of observations is increased. A VAR estimated on data collected over, say, ten years makes use of 120 observations of the monthly variables instead of being limited to the 40 aggregated quarterly observations.

Multiple approaches to dealing with the problem of mixed frequencies are available in the literature. Mixed data sampling (MIDAS) regressions and the MIDAS VAR proposed by \cite{Ghysels2007} and \cite{Ghysels2016}, respectively, use fractional lag polynomials to regress a low-frequency variable on lags of itself as well as high-frequency lags of other variables. This approach is predominantly frequentist, although Bayesian versions are available \citep{Rodriguez2010,Ghysels2016}. A second approach, which is the focus of this work, is to exploit the general ability of state-space modeling to handle missing observations \citep{Harvey1984}. \cite{Eraker2015}, concerned with Bayesian estimation, used this idea to treat intra-quarterly values of quarterly variables as missing data and proposed measurement and state-transition equations for the monthly VAR. \cite{Schorfheide2015} considered forecasting using a construction along the lines of \cite{Carter1994} and provided empirical evidence that the mixed-frequency VAR improved forecasts of eleven US macroeconomic variables as compared to a quarterly VAR. In terms of flexible time-varying models with mixed-frequency data, \cite{Cimadomo2016} employed the mixed-frequency VAR together with time-varying parameters and stochastic volatility to cope with a change in frequency of the data. Following up on the work by \cite{Schorfheide2015}, \cite{Gotz2018} recently showed that more flexible models that include stochastic volatility tend to improve forecasts also within this framework. 

The main contribution of this paper is that we extend the mixed-frequency toolbox by incorporating prior information on the steady states, and by adding stochastic volatility to the model. Thus, we effectively combine the steady-state parametrization of \cite{Villani2009} with the state-space modeling approach for mixed-frequency data of \cite{Schorfheide2015} and the common stochastic volatility model proposed by \cite{Carriero2016}. The proposed model accommodates explicit modeling of the unconditional mean with data measured at different frequencies. In order to employ the model in a realistic forecasting situation, we use a real-time dataset consisting of 13 macroeconomic variables for the US, where ten of the variables are sampled monthly, and the remaining three are available quarterly. We implement the steady-state prior using the standard \cite{Villani2009} approach, and using the hierarchical structure presented by \cite{Louzis2019}. In our empirical application, we find that, for most variables, mixed-frequency data, stochastic volatility, and steady-state information improve forecasting accuracy as compared to models without any of the aforementioned features.

The structure of the paper is as follows. Section \ref{sec:model} describes the main methodology, Section \ref{sec:data} provides information about the data and details about the implementation, and Section \ref{sec:app} evaluates the forecasting performance. Section \ref{sec:con} concludes.

\section{Combining Mixed Frequencies with Steady-State Beliefs}
\label{sec:model}
The mixed-frequency method adopted in this work is a state space-based model which follows the work by \cite{Mariano2010,Schorfheide2015,Eraker2015}. There are several modeling approaches available for handling mixed-frequency data, including MIDAS \citep{Ghysels2007}, bridge equations \citep{Baffigi2004} and factor models \citep{Mariano2003,Giannone2008}. We do not review these further here, but instead refer the reader to the survey by \cite{Foroni2013} and an early comparison conducted by \cite{Kuzin2011}.

\subsection{State-Space Representation of the Mixed-Frequency Model}
To cope with mixed observed frequencies of the data, we assume the system to be evolving at the highest available frequency. This assumption frames the problem of frequency mismatch as a missing data problem. By doing so, the approach naturally lends itself to a state-space representation of the system in which the underlying monthly series of the quarterly variables become the latent states of the system. Because we have a mix of monthly and quarterly frequencies in our empirical application, we will in the following proceed with the presentation of the model from this perspective. It should, however, be noted that other compositions of frequencies are viable within the same framework.

The VAR model at the core of the analysis is specified for the high-frequency and partially missing variables. More specifcally, a VAR($p$) for the $n\times 1$ vector $z_t$ is employed such that
\begin{align}
\Pi(L)z_{t} =  \Phi d_t + u_t, \quad  u_t  \sim \mathrm{N}(0, \Sigma_t),\label{eq:hfvar}
\end{align}
where $\Pi(L)=(I_n-\Pi_1L-\Pi_2L^2-\cdots-\Pi_pL^p)$ is a $p$-th order invertible lag polynomial, $d_t$ is an $m\times 1$ vector of deterministic components and $\Phi$ is an $n\times m$ matrix of parameters. The time index $t$ is here monthly. We let the error term $u_t$ be heteroskedastic and return to the specifics thereof in Section \ref{sec:extending}.

The model in \eqref{eq:hfvar} is a conventional VAR specification, but, in the spirit of \cite{Villani2009}, we instead employ the mean-adjusted form as
\begin{align}
\Pi(L)(z_{t} -\Psi d_t)= u_t, \label{eq:meanadj}
\end{align}
where $\Psi=[\Pi(L)]^{-1}\Phi$. It can be readily confirmed that $\mathrm{E}(z_t|\Pi, \Psi, \Sigma)=\Psi d_t:=\mu_t$, and thus $\mu_t$ is the unconditional mean---\emph{steady state}---of the process. The steady-state representation \eqref{eq:meanadj} requires an explicit prior on the steady state parameters.
However, common practice is to use \eqref{eq:hfvar} with a loose prior on $\Phi$, which implicitly defines an intricate (but loose) prior on $\Psi$ and, subsequently, $\mu_t$. We argue that in many applications, the parametrization in \eqref{eq:meanadj} is more convenient as it allows for a more natural elicitation of prior beliefs. In what follows, we will extend the work of \cite{Villani2009} such that \eqref{eq:meanadj} may still constitute a viable option in the presence of mixed frequencies.

Next, we partition the high-frequency underlying process $z_t$ as $z_{t}=(z_{m, t}', z_{q, t}')'$, where $z_{m,t}$ represents the $n_m$ monthly and $z_{q,t}$ the $n_q$ quarterly variables. Recall that the time $t$ here takes the highest frequency, i.e. monthly. The empirical problem that is ubiquitous in macroeconomic data is that what is observed varies between months such that $z_t$ is not always fully observed.

To distinguish between the underlying process and actual observations, we denote the latter by $y_t$. A consequence of all variables not being observed at every time point $t$ is that the dimension $n_t$ of $y_t$ is not always equal to $n=n_m+n_q$. The observed data in $y_t$ are generally supposed to be some linear aggregate of $Z_t=(z_t', \dots ,z_{t-p+1}')'$ such that
\begin{align}
y_t=\begin{pmatrix}y_{m,t}\\y_{q,t}\end{pmatrix}=\begin{pmatrix}I_{n_m} & 0 \\0 & M_{q, t}\end{pmatrix}\begin{pmatrix}I_{n_m}&0\\0&\Lambda_q\end{pmatrix}Z_t=M_t\Lambda Z_t,\label{eq:y}
\end{align}
where $M_{q, t}$ and $\Lambda_q$ are deterministic selection and aggregation matrices, respectively.

We let $M_{q,t}$ be the $n_q$ identity matrix $I_{n_q}$ if all quarterly variables are observed at time $t$ so that $y_{q, t}=\begin{pmatrix} 0 & \Lambda_{q}\end{pmatrix} Z_t$. In the remaining periods, $M_{q,t}$ is an empty matrix such that $y_t=y_{m,t}$. More complicated observational structures can easily be accomodated using the very same approach; instead of being empty or a full $I_{n}$ matrix, $M_t$ can have rows that correspond to unobserved variables omitted. This idea allows for the approach to seamlessly handle missing data for a subset of the monthly variables at the end of the sample. 

The aggregation matrix $\Lambda_q$ represents the assumed aggregation scheme of unobserved high-frequency latent observations $z_{q, t}$ into occasionally observed low-frequency observations $y_{q, t}$. To make the presentation simpler, we can write the bottom block of $\Lambda Z_t$ as
\begin{align}
\begin{pmatrix}0 & \Lambda_q\end{pmatrix}\begin{pmatrix}z_{m, t} \\z_{q, t}\\ z_{m, t-1}\\ z_{q, t-1}\\ \vdots \\ z_{m, t-p+1}\\ z_{q, t-p+1}\end{pmatrix}=\Lambda_{qq}\begin{pmatrix} z_{q, t} \\ z_{q, t-1} \\ \vdots \\ z_{q,t-p+1}\end{pmatrix},
\end{align}
where $\Lambda_{qq}$ collects the columns of $\Lambda_q$ that correspond to quarterly variables in $Z_t$.

\cite{Schorfheide2015}, working with log-levels of the data, used the intra-quarterly average $y_{q,t}^*=\frac{1}{3}(z_{q, t}^*+z_{q,t-1}^*+z_{q,t-2}^*)$, where $y_{q,t}^*$ denotes the observed quarterly log-levels and $z_{q,t}^*$ the latent monthly log-levels. Because we use log-differenced data, we instead follow \cite{Mariano2003,Mariano2010}. By taking the quarterly difference of $y_{q,t}^*$ to construct our observed growth rates, we obtain
\begin{equation}
\begin{split}
y_{q,t}&=y_{q,t}^*-y_{q,t-3}^* \\
&=\frac{1}{3}\left[(z_{q,t}^*-z_{q,t-3}^*)+(z_{q,t-1}^*-z_{q,t-4}^*)+(z_{q,t-2}^*-z_{q,t-5}^*)\right]\\
&=\frac{1}{3}\left[\left(\Delta z_{q,t}^*+\Delta z_{q,t-1}^*+\Delta z_{q,t-2}^*\right)+\left(\Delta z_{q,t-1}^*+\Delta z_{q,t-2}^*+\Delta z_{q,t-3}^*\right)\right.\\
&\quad+  \left. \left(\Delta z_{q,t-2}^*+\Delta z_{q,t-3}^*+\Delta z_{q,t-4}^*\right) \right],
\end{split}
\end{equation}
Finally, the expression can be written as
\begin{align}
y_{q, t} = \frac{1}{3}\left[\Delta z_{q,t}^*+2\Delta z_{q,t-1}^*+3\Delta z_{q,t-2}^*+2\Delta z_{q,t-3}^*+\Delta z_{q,t-4}^*\right].\label{eq:triangular}
\end{align}
Because the set of weights in \eqref{eq:triangular} sum to three, we define our latent variable of interest to be $z_{q,t}=3 \Delta z_{q,t}^*$, i.e. the latent month-on-month growth rate scaled to be commensurate in scale with the quarterly level.

Equations \eqref{eq:meanadj} and \eqref{eq:y} form a state-space model that can be used for estimation of the model. \cite{Schorfheide2015} suggested an efficient compact formulation of the employed state-space model that is statistically equivalent but computationally more convenient. The compact treatment is based on the observation that the set of monthly variables included in the model are observed for all time points except for a handful at the end of the sample, known as a ragged edge \citep{Banbura2011}. The treatment proposed by \cite{Schorfheide2015} is to let the monthly variables enter the model as exogenous for $t=1, \dots, T_b$, where $T_b$ denotes the final time period where the monthly variables are all observed. By this approach, the monthly variables are excluded from the state equation. The state dimension is thereby reduced from $np$ to $n_q(p+1)$, which improves the computational efficiency substantially. 

In order to more formally introduce this formulation of the model, we first let
\begin{align}\tilde{y}_t=y_t-M_t\Lambda\begin{pmatrix}\Psi d_t \\ \vdots \\ \Psi d_{t-p+1}\end{pmatrix}
\end{align}
denote the mean-adjusted data. The state-space model is thereafter formulated in terms of $\tilde{y}_t$ and $\tilde{z}_t=z_t-\Psi d_t$, leading to the model
\begin{align}
\begin{pmatrix} \tilde{y}_{m, t}\\ \tilde{y}_{q, t}\end{pmatrix} &= \begin{pmatrix}0_{n_m \times n_q} & \Pi_{mq}\\ M_{q,t}\Lambda_q & 0_{n_q\times n_q}\end{pmatrix}\begin{pmatrix}\tilde{z}_{q, t}\\ \tilde{Z}_{q, t-1}\end{pmatrix}+\Pi_{mm}\tilde{Y}_{m,t-1}+\begin{pmatrix}u_{m, t}\\ 0_{n_q\times 1} \end{pmatrix}\\
\begin{pmatrix}\tilde{z}_{q, t}\\ \tilde{Z}_{q, t-1}\end{pmatrix}&=\begin{pmatrix}\Pi_{qq}& 0_{n_q\times n_q}\\ I_{n_qp} & 0_{n_qp\times n_q} \end{pmatrix}\begin{pmatrix}\tilde{z}_{q, t-1}\\ \tilde{Z}_{q,t-2}\end{pmatrix}+\Pi_{qm}\tilde{Y}_{m,t-1}+\begin{pmatrix}u_{q,t} \\ 0_{n_qp\times 1} \end{pmatrix}
\end{align}
where $\Pi_{i,j}$, $i, j\in \{m, q\}$ refer to the submatrices of regression parameters relating the $j$ frequency variables to the conditional mean of the $i$ frequency variables. The errors are the corresponding partitions of $u_t=(u_{m, t}', u_{q,t}')'$ and are consequently correlated. Finally, $\tilde{Y}_{m,t-1}$ stacks the mean-adjusted monthly variables as $\tilde{Y}_{m,t-1}=(\tilde{y}_{m, t-1}', \dots, \tilde{y}_{m, t-p}')'$ and $\tilde{Z}_{q,t-1}=(\tilde{z}_{q, t-1}', \dots, \tilde{z}_{q, t-p}')'$.

The above state-space model remains valid as long as $t\leq T_b$, implying that all of the monthly series are observed. To deal with ragged edges and unbalanced monthly data for $t=T_b+1$, we follow \cite{Ankargren2019} and adaptively add the monthly series with missing data as appropriate. Contrary to \cite{Schorfheide2015}, we thereby avoid use of the full companion form altogether.

\subsection{Extending the Basic Steady-State Model}
\label{sec:extending}
The standard BVAR with the steady-state prior typically produces good forecasts and is for this reason used by e.g. Sveriges Riksbank as one of its main forecasting models (see \citealp{Iversen2016}). However, recent work in the VAR literature demonstrates that allowing for more flexibility may be beneficial. Particularly, letting the error covariance matrix in the model vary over time by incorporating stochastic volatility often improves the predictive ability as demonstrated by e.g. \cite{Clark2011,Clark2013,Carriero2016}. Moreover, studies such as \cite{Banbura2010,Giannone2015,Koop2013} have shown that medium-sized models including 10--20 variables often outperform smaller models when forecasting. The caveat, however, when extending the size of the model under the use of the steady-state prior is that the researcher must set a prior mean and variance for the unconditional mean for each variable in the model. For key variables such as inflation, GDP growth and unemployment this task is relatively effortless, but it can be more challenging when the previous literature does not offer any guidance on reasonable prior specifications. To simplify the process of specifying the steady-state prior, \cite{Louzis2019} developed a hierarchical prior for the steady-state prior that effectively relieves the researcher from eliciting the prior variances of the steady-state parameters. Instead, only prior means are required. Providing a sensible prior for the unconditional mean is generally much simpler than quantifying the uncertainty of one's specification. We next briefly describe the stochastic volatility and hierarchical steady-state prior specifications that we extend our basic model with.

\subsubsection{Stochastic volatility}
The stochastic volatility model we employ is the common stochastic volatility (CSV) model of \cite{Carriero2016}, which is a parsimonious and simple approach for letting the error covariance matrix in the model vary over time. The state equation describing the high-frequency VAR is under the CSV variance specification given by 
\begin{align}
\Pi(L)(z_t-\Psi d_t)&=\sqrt{f_t}A^{-1}e_t, \quad e_t \sim \mathrm{N}(0, I)\label{eq:model}
\end{align}
where $A^{-1}$ is a lower triangular matrix and $f_t$ is the latent univariate volatility series evolving according to
\begin{align}
 \log(f_t)&=\phi \log(f_{t-1})+\nu_t, \quad \nu_t \sim \mathrm{N}(0, \sigma^2).\label{eq:factor}
\end{align}
The log-volatility $\log(f_t)$ thus evolves as an AR(1) process without intercept with parameters ($\phi, \sigma^2$). The time-varying error covariance matrix implied by the preceding model is $\Sigma_t=f_t\Sigma$, where $\Sigma=A^{-1}(A^{-1})'$. Consequently, the CSV prior assumes a fixed covariance structure where the volatility factor provides a time-varying scaling of the constant error covariance $\Sigma$.

\subsubsection{Hierarchical steady-state priors}
The appealing feature of the steady-state prior is that it allows the researcher to use readily available information about long-run steady-state levels of the included variables. For the reasons discussed earlier, \cite{Louzis2019} proposed a hierarchical steady-state prior using the normal-gamma construction used by e.g. \cite{Griffin2010,Huber2017}. The reason for such an approach is that the benefits of the steady-state prior are larger when we have accurate and relatively informative priors for the steady states. The normal-gamma prior employs a hierarchical specification that provides sufficiently heavy tails to allow for a large degree of shrinkage to the prior mean when appropriate, and more flexibility otherwise. In effect, the researcher only has to provide a prior mean for each steady-state parameter as the associated variances are instead obtained from the hyperparameters higher up in the hierarchy.

To be more precise, the hierarchical steady-state prior is based on the normal-gamma prior proposed by \cite{Griffin2010} that employs a hierarchical specification given by
\begin{align}
\psi_j|\omega_{\psi, j} &\sim \mathrm{N}(\underline{\mu}_{\psi,j}, \omega_{\psi, j})\\
\omega_{\psi, j} &\sim \mathrm{G}(\phi_\psi, 0.5\phi_\psi\lambda_\psi),
\end{align}
where $\phi_\psi$ and $\lambda_\psi$ are additional fixed hyperparameters and $\mathrm{G}(a, b)$ denotes the gamma distribution with shape $a$ and rate $b$. The prior is therefore constructed using idiosyncratic, or local, hyperparameters $\omega_{\psi, j}$, which in turn depend on the two auxiliary hyperparameters $\phi_\psi$ and $\lambda_\psi$. 

\cite{Griffin2010} showed that the variance of the unconditional prior for $\psi_j$ is negatively associated with $\lambda_\psi$, meaning that higher values of $\lambda_\psi$ induce a larger degree of shrinkage towards the prior mean. The hyperparameter $\lambda_\psi$ can therefore be interpreted as a global shrinkage parameter. At the same time, the excess kurtosis of the unconditional prior is negatively related to $\phi_\psi$. Taken together, the implication is that if a tight prior (i.e. $\lambda_\psi$ is high) is employed, the local shrinkage given by $\omega_{\psi, j}$ can still deviate notably from zero if $\phi_\psi$ is small due to the heavy tails of the unconditional prior distribution. This feature allows for a shrinkage profile that is in general tight, but loose when necessary.

\subsection{Prior Distributions}

We use a standard normal inverse Wishart prior for the VAR coefficients and error covariance $(\Pi, \Sigma)$. Thus, we have a priori
\begin{equation}
\Sigma \sim \mathrm{IW}(\underline{S}, \underline{\nu}), \quad  \vc(\Pi')| \Sigma \sim \mathrm{N}(\vc(\underline{\Pi}'), \Sigma \otimes  \underline{\Omega}_{\Pi}),
\end{equation}
where $\Pi=(\Pi_1, \dots, \Pi_p)$. The main diagonal of the prior covariance matrix for the regression parameters, $\underline{\Omega}_{\Pi}$, is set in the Minnesota-style fashion
\begin{align}
\underline{\omega}_{\Pi, ii} =
		\frac{\lambda_1^2}{(l^{\lambda_2}s_r)^2}  & \mbox{ for lag \textit{l} of variable \textit{r}},\: i = (l-1)p+ r
\end{align}
where $\lambda_1$ is the overall tightness and $\lambda_2$ determines the lag decay rate; the inclusion of $s_r$ adjusts for differences in measurement scale of the variables. For a more thorough exposition of the normal inverse Wishart prior, the reader is referred to \cite{Karlsson2013}.

While $\Sigma$ describes the fixed covariance structure, the time-varying volatility in the model is governed by the latent volatility $f_t$. For the two parameters associated with its evolution, $(\phi, \sigma^2)$, we use a normal distribution truncated to the stationary region for $\phi$, and an inverse gamma prior for $\sigma^2$:
\begin{align}
\phi &\sim \mathrm{N}(\underline{\mu}_\phi, \underline{\Omega}_{\phi}; \, |\phi|<1)\\
\sigma^2 &\sim \mathrm{IG}(\underline{d}\cdot\underline{\sigma}^2, \, \underline{d}).
\end{align}

As discussed in Section \ref{sec:extending}, the priors for the steady-state parameters are normal conditional on the local shrinkage parameters. Instead of fixing the top-level hyperparameters $\phi_\psi$ and $\lambda_\psi$, \cite{Huber2017} proceeded with an additional hierarchy by specifying priors for $\phi_\psi$ and $\lambda_\psi$. We follow their suggestion and obtain the following hierarchical prior specification for the steady-state parameters:
\begin{align}
\psi_j|\omega_{\psi, j} &\sim \mathrm{N}(\underline{\mu}_{\psi, j}, \omega_{\psi, j})\\
\omega_{\psi, j}|\phi_{\psi}, \lambda_\psi &\sim \mathrm{G}(\phi_\psi, 0.5\phi_\psi\lambda_\psi)\\
\phi_\psi&\sim \mathrm{Exp}(1) \\
\lambda_\psi&\sim \mathrm{G}(c_0,c_1).
\end{align}

\subsection{Posterior Sampling}
To estimate the model and produce forecasts, we employ Markov Chain Monte Carlo (MCMC). The MCMC algorithm consists of multiple Gibbs sampling steps, which we describe next. We relegate some of the details to Appendix \ref{sec:posterior}.
\paragraph{Sampling the latent monthly variables}
To sample from the posterior distribution of the latent monthly variables, $p(Z|\Pi, \Sigma, \psi, f, Y, d)$, we use a simulation smoother along the lines of \cite{Durbin2012}. To increase the computational efficiency, we implement it using the compact formulation for the balanced part of the sample as suggested by \cite{Schorfheide2015}. For the unbalanced ragged edge, we instead leverage the adaptive procedure developed by \cite{Ankargren2019}. The simulation smoothing step is conducted based on the mean-adjusted data $\tilde{y}_t$ to produce a draw of $\tilde{z}_t$. We thereafter construct the unadjusted high-frequency series by adding the deterministic component $z_t=\tilde{z}_t+\Psi d_t$.

\paragraph{Sampling the regression and covariance parameters}
Given $Z, \psi$ and $f$, the VAR can be transformed into a homoskedastic VAR without intercept based on $\bar{z}_t=\tilde{z}_t/\sqrt{f}_t$ and $\bar{Z}_{t-1}=(\bar{z}_{t-1}', \dots, \bar{z}_{t-p}')'$:
\begin{align}
\bar{z}_t=\Pi \bar{Z}_{t-1}+A^{-1}e_t.\label{eq:zbar}
\end{align}
By standard results \citep{Kadiyala1993,Kadiyala1997}, the conditional posterior distribution is also normal inverse Wishart. It is thereby possible to sample from the marginal posterior of $\Sigma$ followed by the full conditional posterior of $\Pi$:
\begin{align}
\Sigma|\bar{Z}& \sim \mathrm{IW}(\overline{S}, T + \nu)\\
\vc(\Pi')|\Sigma, \bar{Z}&\sim \mathrm{N}\left(\vc(\bar{\Pi}'), \Sigma \otimes \bar{\Omega}_{\Pi}\right).
\end{align}

The posterior moments are standard given the transformation of the model and presented in Appendix \ref{sec:posterior}.
A draw can efficiently be made from the posterior of $\Pi$ by reverting to its matrix-normal form:
\begin{align}
\Pi=\mathrm{chol}(\bar{\Omega}_\Pi^{-1})'\setminus \left[\mathrm{chol}(\bar{\Omega}_\Pi^{-1})\setminus\left(\underline{\Pi}\,\underline{\Omega}_\Pi+\sum_{t=1}^T \bar{z}_t\bar{Z}_{t-1}'\right)+\Xi \times\mathrm{chol}(\Sigma)'\right],
\end{align}
where $\Xi$ is an $n\times np$ matrix of numbers independently drawn from the standard normal distribution, $\mathrm{chol}$ is the lower triangular Cholesky decomposition and the operation $A\setminus B$ means to solve the linear system $A X= B$ for $X$. Because the Cholesky factor is triangular, the linear systems can be solved more efficiently using forward and back substitution.

\paragraph{Sampling the steady-state parameters}
Prior to sampling the steady-state parameters, the associated hyperparameters are drawn from their respective conditional posterior distributions. The conditional posterior of the global shrinkage parameter $\lambda_\psi$ is gamma distributed and given by
\begin{align}
\lambda_\psi &\sim \mathrm{G}\left(nm\phi_\psi+c_0, \, 0.5\phi_\psi\sum_{j=1}^{nm}\omega_{\psi, j}+c_1\right).
\end{align}
The conditional posterior of $\phi_\psi$ is proportional to
\begin{align}
p(\phi_\psi|\omega_\psi, \lambda_\psi)&\propto g(\phi_\psi|\omega_\psi, \lambda_\psi)\\
&=\prod_{j=1}^{nm}\frac{(0.5\lambda_\psi\phi_\psi)^{\phi_\psi}}{\Gamma(\phi_\psi)}\omega_{\psi, j}^{\phi_\psi-1}\exp\left\{-0.5\lambda_\psi\phi_\psi\omega_{\mu, j}-\phi_\psi\right\}
\end{align}
and permits no representation in terms of a standard distribution. As suggested by \cite{Huber2017,Louzis2019} we employ a random walk Metropolis-Hastings step in order to sample from the posterior distribution. The random walk operates on the log-scale and the proposal is given by 
\begin{align}
\log(\phi_\psi^{*})=\log(\phi_\psi^{(i-1)})+s z, \quad z \sim \mathrm{N}(0, 1),
\end{align}
where $s$ is a scaling factor. The proposed value $\phi_\psi^*$ is accepted with probability 
\begin{align}
r = \min\left\{1, \frac{g(\phi_\psi^*|\omega_\psi, \lambda_\psi)}{g(\phi_\psi|\omega_\psi, \lambda_\psi)}\frac{\phi_\psi^*}{\phi_\psi^{(i-1)}}\right\},
\end{align}
where the second ratio accounts for the asymmetric proposal distribution.

Given the hyperparameters, the local shrinkage parameters $\omega_{\psi, j}$ can be sampled. The conditional posterior distribution is the generalized inverse Gaussian distribution
\begin{align}
\omega_{\psi, j}|\lambda_\psi, \phi_\psi, \psi_j &\sim \mathrm{GIG}\left(\phi_\psi-0.5, \, \lambda_\psi\phi_\psi, \, (\psi_j-\underline{\mu}_{\psi, j})^2\right) \quad j=1, \dots, nm,
\end{align}
where if $y\sim \mathrm{GIG}(a, b, c)$ then $p(y; a, b, c)\propto y^{a-1}\exp\left\{0.5(by+c/y)\right\}$. The prior covariance matrix for $\psi$, i.e. $\underline{\Omega}_\psi$, can thereafter be constructed as the diagonal matrix with main diagonal given by $(\omega_{\psi, 1}, \dots, \omega_{\psi, nm})$.

Next, by dividing both sides of the model \eqref{eq:model} by $\sqrt{f}_t$ we obtain a homoskedastic model given by
\begin{align}
\Pi(L)\left(\frac{z_t}{\sqrt{f}_t}-\Psi\frac{d_t}{\sqrt{f}_t}\right)=A^{-1}e_t.
\end{align}
The posterior moments provided by \cite{Villani2009} therefore apply directly for the preceding transformation of the model. Let 
\begin{align}
\check{z}_t=\Pi(L)z_t/\sqrt{f}_t, &\quad\check{d}_t'=\begin{pmatrix}\frac{d_t'}{\sqrt{f}_t}& -\frac{d_{t-1}'}{\sqrt{f}_t} & \cdots & \frac{d_{t-p}'}{\sqrt{f}_t}\end{pmatrix}\\
U&=\begin{pmatrix}I_{nm} \\ I_m \otimes \Pi_1 \\\vdots \\I_m \otimes \Pi_p\end{pmatrix}.
\end{align}
The posterior distribution of $\psi$ is
\begin{align}
\psi|\check{Z}, \check{d}, \omega_\psi \sim \mathrm{N}(\overline{\mu}_{\psi}, \overline{\Omega}_\psi)
\end{align}
with posterior moments
\begin{align}
\overline{\Omega}_\psi^{-1} &= \underline{\Omega}_\psi^{-1}+ U'\left[\left(\sum_{t=1}^T\check{d}_t\check{d}_t'\right)\otimes \Sigma^{-1}\right]U\\
\overline{\mu}_\psi &= \overline{\Omega}_\psi \left[U'\vc\left(\Sigma^{-1}\sum_{t=1}^T \check{z}_t\check{d}_t'\right)+\underline{\Omega}_\psi^{-1}\underline{\psi}\right].
\end{align}
\paragraph{Sampling the latent volatility}
Conditional on the other parameters in the model, we can obtain
\begin{align}
\ddot{z}_t=A\Pi(L)(z_t-\Psi d_t)=\sqrt{f_t}e_t.
\end{align}
Squaring and taking the logarithm of the elements of $\ddot{z}_t$ yields
\begin{align}
\log(\ddot{z}_{i,t}^2)=\log(f_t)+\log(e_{i,t}^2), \quad i=1,\dots,n,
\end{align}
where $\ddot{z}_{i,t}$ is the $i$th element of $\ddot{z}_t$ with a similar logic for $e_{i,t}$.
Coupling the preceding equation with the transition equation \eqref{eq:factor} defines a linear but non-normal state-space model. \cite{Kim1998} proposed a sampling strategy that introduces auxiliary mixture indicators $r_{t,i}$ so that the model conditional on these indicators is normal. We use the refined ten-state mixture by \cite{Omori2007} together with the algorithm discussed by \cite{McCausland2011} and as implemented by \cite{Kastner2014} to sample from the posterior distribution of the latent volatility series. 

The posteriors of the parameters of the volatility process are standard given $f$. The posterior distribution of $\phi$ is a truncated normal distribution whereas the posterior distribution of $\sigma^2$ is inverse gamma. We proceed by sampling $(\phi, \sigma^2)$ first, the mixture indicators $r_{t,i}$ next and, finally, the latent volatility series in order to target the correct posterior distribution as discussed by \cite{DelNegro2015}.

\section{Data and Implementation Details}
\label{sec:data}
In this section, we provide information about the data used and some details regarding the implementations.
\subsection{Data}
Our dataset consists of 13 key macroeconomic variables for the United States. The dataset we use largely parallels that of \cite{Carriero2016,Louzis2019} with the exception that we use CPI inflation as the sole measure of inflation. The data consist of ten monthly and three quarterly variables and ranges over the period 1980M01--2018M12. Most of the included variables are available with real-time vintages in the ALFRED database. For variables not available in ALFRED, we turn to FRED and FRED-MD \citep{McCracken2016}. A summary of the data is provided in Table \ref{tab:data}.

\begin{table}
\footnotesize
\centering
\caption{Summary of the Real-Time Dataset}
\label{tab:data}
\begin{threeparttable}
\begin{tabular}{llllll}
\toprule
Series  & Transformation  & Frequency & Real time & $\underline{\mu}_{\psi, j}$ & $\sqrt{\underline{\omega}_{\psi, j}}$\\
\midrule
Nonfarm payrolls$^\ast$ & $1200\Delta \ln$  & Monthly & Yes & 3 & 0.5\\
Hours$^{\ast\dagger}$ & X13, $1200\Delta \ln$  & Monthly & $\geq 2011$ & 3 & 0.5 \\
Unemployment rate$^\ast$ & None & Monthly & Yes & 6 & 1\\
Federal funds rate$^\ast$ & None  & Monthly & Yes & 5 & 0.7\\\vspace{0.5em}
Bond spread$^\dagger$ & Monthly ave. & Monthly & Yes & 1 & 1\\
Stock market index$^\ddagger$ & $1200\Delta \ln$ & Monthly & No & 0 & 2 \\
Personal consumption$^\ast$ & $1200\Delta \ln$ & Monthly & Yes & 3 & 0.7 \\
Industrial production$^\ast$ & $1200\Delta \ln$  & Monthly & Yes & 3 & 0.7 \\
Capacity utilization$^\ast$ & None  & Monthly & Yes &80 & 0.7 \\\vspace{0.5em}
CPI inflation$^\ast$ & $1200\Delta \ln$  & Monthly & Yes & 2 & 0.5\\
Nonresidential inv.$^\ast$ & $400\Delta \ln$ & Quarterly & Yes & 3 & 1.5\\
Residential inv.$^\ast$  & $400\Delta \ln$  & Quarterly & Yes & 3 & 1.5\\
GDP growth$^\ast$  & $400\Delta \ln$  & Quarterly & Yes & 2 & 0.5\\
\bottomrule
\end{tabular}
{\footnotesize
\emph{Sources:}
\begin{tablenotes}
    \item[$\ast$] ALFRED, Federal Reserve Bank of St. Louis
    \item[$\dagger$] FRED, Federal Reserve Bank of St. Louis
    \item[$\ddagger$] FRED-MD, \cite{McCracken2016}
      \end{tablenotes}
}
{\footnotesize
\emph{Notes:}
\begin{enumerate}[noitemsep,topsep=0pt]
\item Real-time data for Hours is available in ALFRED from 2011 and onwards; data from FRED is used prior to 2011. Hours is seasonally adjusted using X-13ARIMA-SEATS using the \texttt{seasonal} package in R \citep{Sax2018}. 
\item A list of the IDs of the variables is available in Appendix \ref{app:data}.
\end{enumerate}
}
\end{threeparttable}
\end{table}

We follow \cite{Louzis2019} and transform the raw series to growth levels. For our monthly variables, we use month-on-month growth rates, whereas the three quarterly variables are computed as quarter-on-quarter rates. All growth rates are annualized. The final two columns of Table \ref{tab:data}, $\underline{\mu}_{\psi,j}$ and $\sqrt{\underline{\omega}}_{\psi, j}$, display the prior means and prior standard deviations of the unconditional means of the variables. The values are drawn from \cite{Louzis2019}, but are also in line with e.g. \cite{Clark2011,Osterholm2012}.

\begin{figure}
\centering
\iffig
\includegraphics[width=\textwidth]{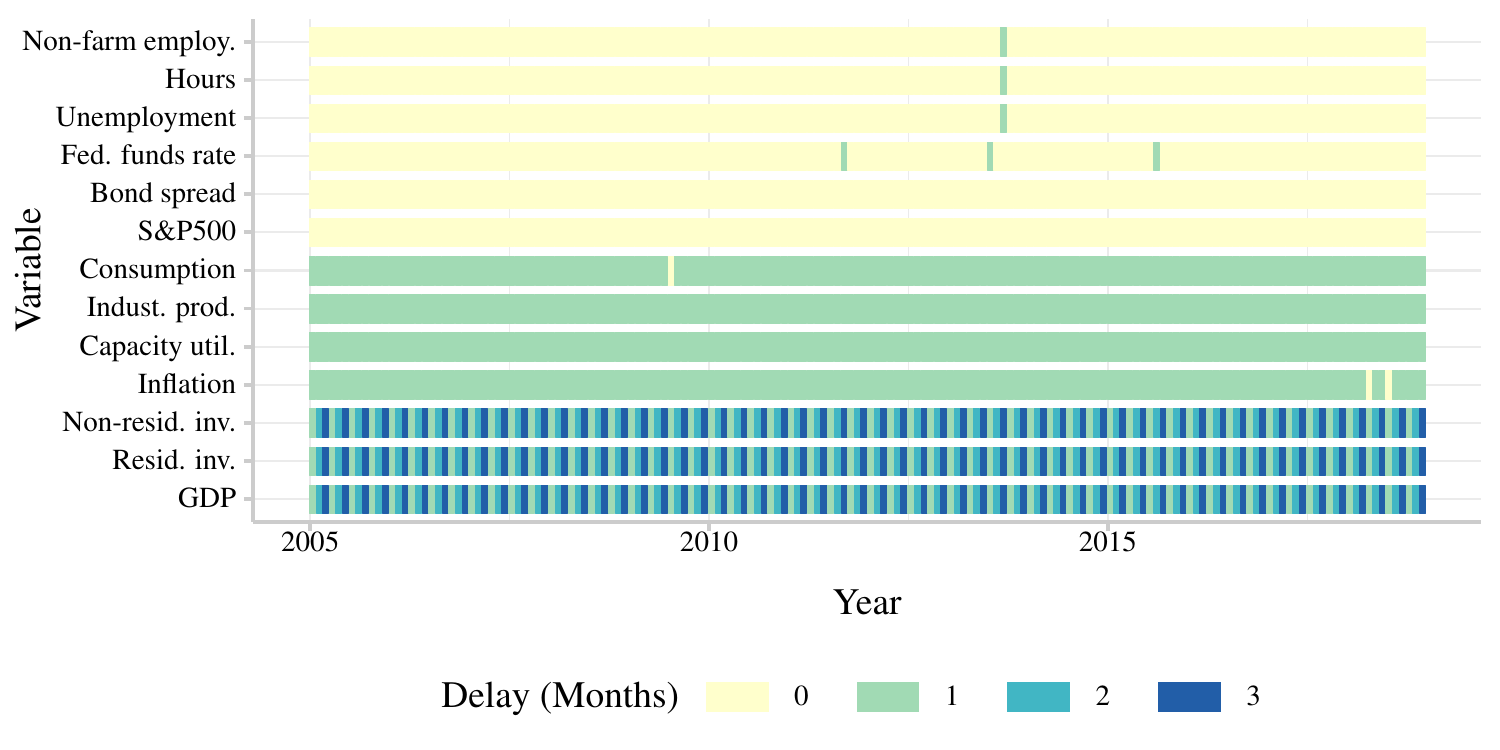}
    \caption{Publication Delays. The shade of each box represents the number of months since the last available observation. The delay is computed for the tenth day of the corresponding month; a zero-month delay implies that the observation for the preceding month is available.}
\else
\caption{Around here}
\fi
    \label{fig:pubdel}
\end{figure}

We use real-time data where available throughout the forecasting exercise. To obtain a realistic pattern of available observations, we first consider the information set available on the tenth day of every month. Figure \ref{fig:pubdel} displays the publication pattern during 2005--2018 and shows the number of months that has passed since the last available publication.

Figure \ref{fig:pubdel} shows a pattern that is characteristic of real-time forecasting of macroeconomic data. Data for financial and select real and nominal variables are already available for the previous month, whereas the previous month's outcomes for some of the monthly variables are unknown.  The pattern of availability displayed shows that consumption and inflation are available with a one-month delay at every month except for a handful of occasions. Similarly, non-farm employment, hours, unemployment and the federal funds rate are typically available with a zero- month delay with the exception of a few months. In the final dataset that we use in our forecasting exercise, we make adjustments to the publication delays in order to obtain a more uniform dataset. The adjustments change the publication structure in the vintages so that the aforementioned variables have the same delay in all vintages, i.e. consumption and inflation are always observed wtih a delay of one month, whereas non-farm employment, hours, unemployment and the federal funds rate are always observed without any delay. Consequently, at every month that we make our forecasts observations are available for the preceding month for six of the monthly variables, whereas four still lack data. 

\subsection{Implementation Details}
The mixed-frequency models that we estimate use $p=12$ lags following e.g. \cite{Banbura2010}. The overall tightness in the prior distribution for the regression parameters is set to $\lambda_1=0.2$ and the lag decay used is $\lambda_2=1$. We use 15,000 draws in the MCMC procedure and discard the first 5,000.

For the hierarchical steady-state prior, we let $c_0=c_1=0.01$ in line with \cite{Huber2017,Louzis2019}. To set the scale of the proposal distribution for $\phi_\psi$, we employ the adaptive scaling procedure discussed by \cite{Roberts2009}. We use a batch size of 100 and check every 100 iterations if the fraction of acceptances within the most recent batch exceeds 0.44. If it does, we increase $s$ by $\delta(k)=\min(0.01, k^{-1/2})$, where $k$ denotes the batch number. If the fraction of acceptances was less than 0.44, $s$ is instead decreased by $\delta(k)$.

For the parameters of the log-volatility process, we let the prior mean and standard deviation for $\phi$ be $\underline{\mu}_\phi=0.9$ and $\sqrt{\underline{\Omega}_\phi}=0.1$, respectively. The prior mean and degrees of freedom of $\sigma^2$ are $\underline{\sigma}^2=0.01$ and $\underline{d}=4$.

\section{Empirical Application: Real-Time Forecasting of Key US Variables}
\label{sec:app}

In this section, we assess the forecasting ability of the model that we propose. The assessment is carried out by studying the out-of-sample predictive accuracy of the model based on the real-time dataset for the US that was discussed in Section \ref{sec:data}.

\subsection{Forecasting Setup}

The quarterly steady-state Bayesian VAR model has been used in several previous studies, see for example \cite{Adolfson2007b,Osterholm2008,Villani2009,Clark2011,Ankargren2017}. The model is employed both for policy purposes and for forecasting and is implemented in the Matlab toolbox BEAR developed at the European Central Bank \citep{Dieppe2016}. Our empirical application targets this audience, and our main interest lies in seeing whether the components we add to the model---mixed frequencies, stochastic volatility and hierarchical steady states---improve upon the benchmark model of \cite{Villani2009} estimated on single-frequency data. The forecasting results are also compared to models using Minnesota-style normal inverse Wishart priors, i.e. without use of the steady-state component. A summary of the models that we include in the forecast evaluation is presented in Table \ref{tab:models}.

\begin{table}
\small
\caption{List of Models}\centering
\label{tab:models}
\begin{tabularx}{\textwidth}{lX }
\toprule
Model & Description \\
\midrule
Benchmark & Single-frequency model with the steady-state prior and a normal inverse Wishart prior for ($\Pi, \Sigma$), constant error covariance. Includes all 13 variables aggregated to the quarterly frequency or the ten monthly variables depending on context.\vspace{1em}\\
Minn-IW & Normal-inverse Wishart prior, constant error covariance\\
Minn-CSV & Normal-inverse Wishart prior with common stochastic volatility\vspace{0.5em}\\
SS-IW & Steady-state prior with a normal inverse Wishart prior for ($\Pi, \Sigma$), constant error covariance\\
SS-CSV & Steady-state prior with a normal inverse Wishart prior for ($\Pi, \Sigma$) with common stochastic volatility\vspace{0.5em}\\
SSNG-IW & Hierarchical normal-gamma steady-state prior with a normal inverse Wishart prior for ($\Pi, \Sigma$), constant error covariance\\
SSNG-CSV & Hierarchical normal-gamma steady-state prior with a normal inverse Wishart prior for ($\Pi, \Sigma$) with common stochastic volatility\\
\bottomrule
\end{tabularx}
\end{table}

The benchmark model is the steady-state model estimated on single-frequency data. Depending on whether it serves as benchmark for quarterly or monthly variables, we include either the full set of variables (aggregated to the quarterly frequency) or the ten monthly variables. The quarterly VAR uses $p=4$, whereas for the monthly VAR $p=12$.

We use a recursive forecasting scheme to evaluate the forecasting performance of the considered models. Beginning in January 2005, we estimate the models and make forecasts and then recursively add months to the set of data used for estimation. The benchmark models use the balanced data, whereas the mixed-frequency models automatically handle the ragged edges. 

The forecasting ability of the models is evaluated with respect to both point and density forecasts. For point forecasts, we consider the root mean squared errors. For density forecasts, we compute univariate and multivariate log predictive density scores. We do so by fitting a normal density to the draws from the predictive distribution following e.g. \cite{Adolfson2007b,Carriero2015}. That is, we compute
\begin{align}
LPDS_{h, t}^{(m, s)} &=\\
n_s\ln(2\pi)&+\ln\left|V^{(m, s)}_{t+h|t}\right|+(y^{(s)}_{t+h}-\bar{y}^{(m, s)}_{t+h|t})'\left(V^{(m, s)}_{t+h|t}\right)^{-1}(y^{(s)}_{t+h}-\bar{y}^{(m, s)}_{t+h|t}),
\end{align}
where $m$ denotes the model, $s$ denotes the set of variables the LPDS is computed for, $n_s$ is the dimension of $s$, $h$ is the forecast horizon, and $\bar{y}$ and $V$ are the mean and covariance of the draws from the relevant predictive distribution. For fixed $(m, s, h)$, we compute the summary LPDS by averaging over the evaluation period. We calculate the LPDS jointly but separately for the monthly and quarterly variables, and univariately for all variables.

An important question when using real-time data is with respect to what vintage the forecasts should be evaluated. There is no consensus, but two alternatives are more common in the literature. The first, as used by e.g. \cite{Romer2000} and \cite{Clark2011}, is to use the second available vintage. This choice can be justified by acknowledging that revisions that occur after longer periods of time may be unforeseeable and more structural in nature by relating to e.g. definitions, methods of measurement, etc. The second available estimate therefore provides a less noisy estimate than the initial available value, yet is produced in the same environment as the forecaster is active. The second common approach for evaluation, as followed by e.g. \cite{Schorfheide2015}, is to use the most recent vintage. For whatever reason revisions may have taken place, the currently available data provide the best estimates of e.g. inflation and output in previous years. We follow the latter approach and use the most recent vintage for evaluating the forecasts, but for transparency provide the main results of the evaluation using the second available vintage in Appendix \ref{app:second}.

\subsection{In-Sample Estimation}
\begin{figure}
\centering
\iffig
\includegraphics[width=\textwidth]{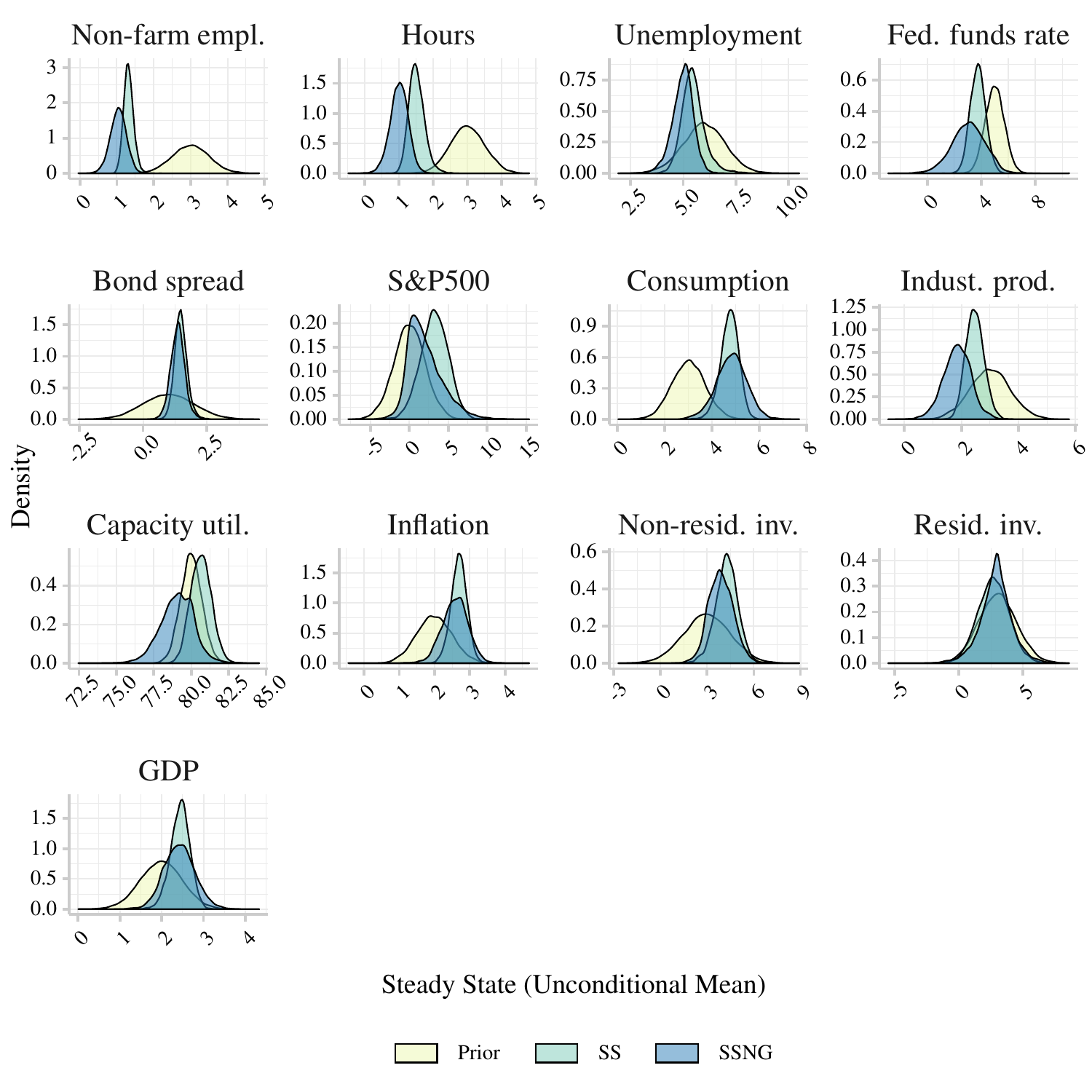}
    \caption{Steady States. Kernel density estimates of the prior and posterior distributions of the steady states (unconditional means).}
\else
\caption{Around here}
\fi
    \label{fig:steadystates}
\end{figure}

As a preliminary analysis, we begin by estimating the mixed-frequency VAR model using the SS and SSNG priors to see whether the obtained steady-state posteriors differ. Because the long-term forecasts are largely determined by the steady-state posterior, seeing whether differences are present is of direct importance for forecasts beyond the immediate short term. Figure \ref{fig:steadystates} displays kernel density estimates of the posterior distributions from the mixed-frequency model with common stochastic volatility. As a point of reference, the figure includes the prior distribution detailed in Table \ref{tab:data}.

As expected, the posteriors in Figure \ref{fig:steadystates} are for the most part similar. The modes of the posteriors are close to perfectly aligned for variables such as bond spread, inflation, residential investment and GDP. For others---e.g., hours, the federal funds rate and industrial production---the SSNG posteriors deviate more from both the priors and the SS posteriors. 

\begin{figure}
\centering
\iffig
\includegraphics[width=\textwidth]{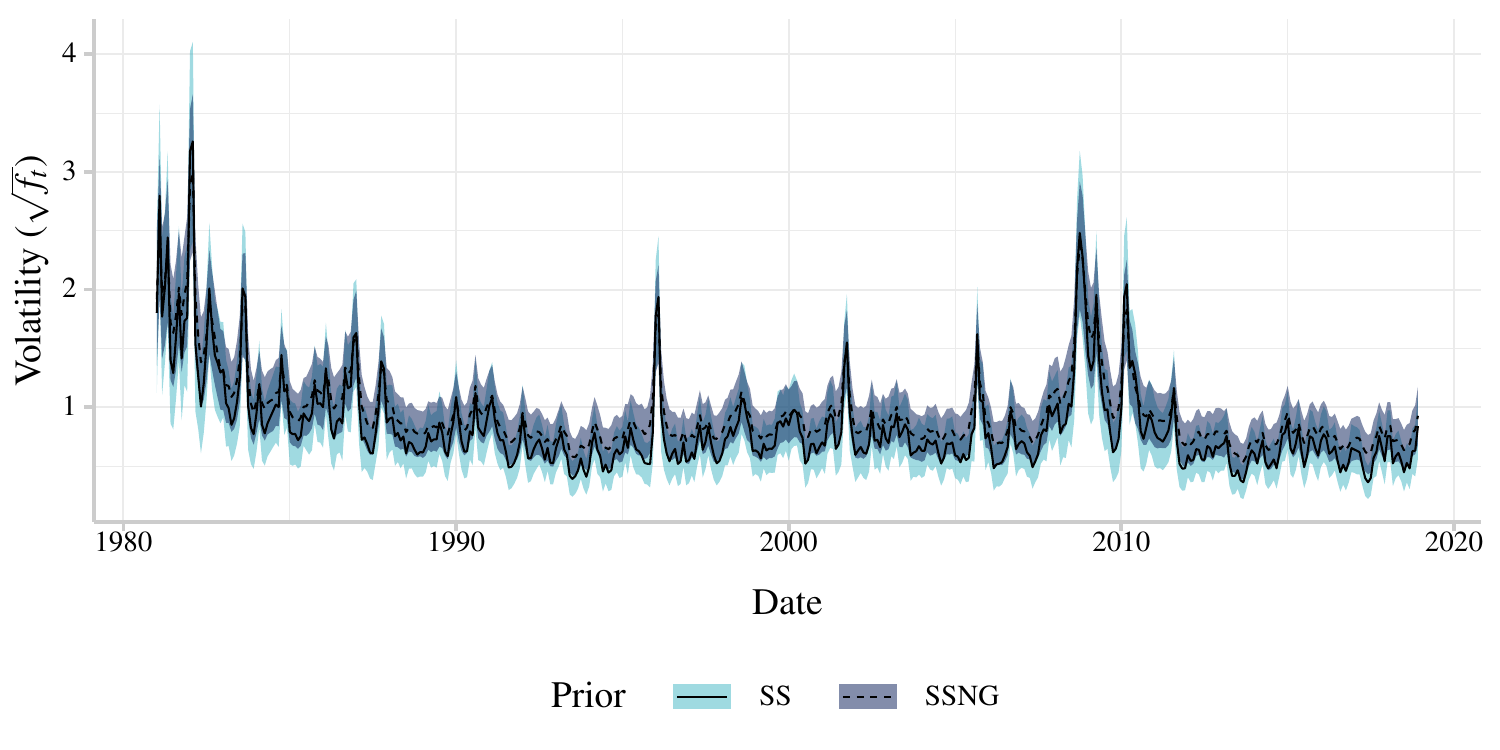}
     \caption{Volatility. The lines display the posterior means of $\sqrt{f}_t$ and the bands show the 90 \% posterior intervals.}
\else
\caption{Around here}
\fi
    \label{fig:volatility}
\end{figure}

While the steady states are of central importance for the levels of the forecasts, the precision thereof is highly influenced by the common stochastic volatility factor. Figure \ref{fig:volatility} displays the mean of $\sqrt{f}_t$ together with 90 \% bands for the SS-CSV and SSNG-CSV models.

Figure \ref{fig:volatility} shows that there is little difference between the estimated volatility factors in the two steady-state models.  Peaks of volatility are aligned and reach the same levels, while the level of the factor in the SSNG model is slightly higher in normal times. Both display the entrance into the Great Moderation in the beginning of the 1980s with heightened volatility again around the recent financial crisis. The interpretation of the level of the factor is that the time-invariant elements in the error covariance matrix $\Sigma$ have been scaled by $f_t$, which roughly amounts to an amplification by a factor of 4--6 during the recent financial crisis and a compression of around 0.5--0.75 in recent years. This feature has a direct effect on the width of the predictive distribution.

\subsection{Forecast Evaluation}
In this section, we present the main results of the forecast evaluation. For space considerations, the presentation includes results from the joint evaluations as well as univariate results for the three quarterly variables and the three monthly variables that are typically of primary interest: the inflation, federal funds and unemployment rates. For completeness, univariate evaluation results for the remaining variables can be found in Appendix \ref{sec:addres}.

\paragraph{Joint forecasting results}

Table \ref{tab:lpds} presents the results from the LPDS computed jointly. We compute the LPDS separately for the set of quarterly and monthly variables, respectively. The forecast horizons $h$ in the table correspond to the frequency of the respective set of variables.

Across all horizons and sets of variables, SS-CSV and SSNG-CSV dominate with only one exception in which Minn-CSV does slightly better than SS-CSV. For the quarterly sets of variables, SS-CSV outperforms the other models for $h>0$ with the SSNG-CSV model ranking first for the nowcast. Minn-CSV ranks higher than the constant volatility models for the initial horizons, but for the long-term forecasts the added value of the steady-state prior outweighs the improvements obtained from stochastic volatilities. However, given a model, stochastic volatility appears to be useful as it improves the joint forecasting performance of quarterly variables across the board when comparing the constant volatility models to their heteroskedastic counterparts. Within the two groups of models with constant and stochastic volatility, we see that the steady-state models forecast better than Minn-IW and Minn-CSV, respectively, throughout all horizons. Therefore, the table shows that steady-state information and flexible modeling of the volatility structure help to improve the quarterly forecasts.

For the performance of the monthly forecasts, the picture is largely the same. The three models with stochastic volatility outperform the constant models for all horizons and SSNG-CSV produces the most accurate density forecasts for $h=2,3,4$. For the remaining horizons, SS-CSV picks up the lead. Among the constant volatility models, the ranking is no longer uniform across horizons.

With respect to the joint log predictive scores, we can therefore conclude the following. First, there are gains in utilizing prior information on the steady states. Second, further improvements can be obtained by allowing for stochastic volatility. Third, with a handful of exceptions for the quarterly forecasts made by Minn-IW and SS-IW, the relative LPDS is negative throughout, indicating that the mixed-frequency models produce better density forecasts than the single-frequency benchmarks. The three points are in line with the previous literature and can be seen as a synthesis of the conclusions made by \cite{Villani2009,Clark2011,Schorfheide2015,Carriero2016,Louzis2019}.

\paragraph{Quarterly univariate forecasting results}
Tables \ref{tab:gdpc1}--\ref{tab:pnfi} present the univariate LPDS and RMSE for the three quarterly variables GDP, Residential investment and Non-residential investment. 

Starting with GDP, a somewhat different pattern than what was seen for the joint LPDS emerges. For both evaluation metrics, SS-IW is generally the better forecaster beyond the short term and is only outperformed by CSV models at the first three horizons and in terms of density forecasts. Table \ref{tab:gdpc1} shows that the mixed-frequency models do better than the quarterly benchmark for the immediate short term when either nowcasting the current quarter or forecasting the next quarter. Beyond the first quarter forecast, the quarterly model generally produces more accurate forecasts. A similar result is found by \cite{Schorfheide2015}. Use of the steady-state prior results in more accurate forecasts at every horizon, but whether or not a hierarchical prior formulation and stochastic volatility provide improvements varies. The homoskedastic steady-state models outperform the Minn-IW model at all horizons, and the stochastic volatility steady-state models consistently forecast GDP growth more accurately than Minn-CSV.

For residential investment, Table \ref{tab:prfi} presents forecasting results that more closely resemble the joint results. SS-CSV and SSNG-CSV dominate for all horizons, although the difference with respect to Minn-CSV is occasionally small, particularly for the point forecasts. Nevertheless, both steady-state models with stochastic volatility perform well with better scores than all other models for every horizon and with respect to both point and density forecasts.

Finally, Table \ref{tab:pnfi} shows the forecast evaluation for Non-residential investment. The pattern displayed in Table \ref{tab:pnfi} is a mix of the patterns in Tables \ref{tab:gdpc1}--\ref{tab:prfi}. For the nowcast, Minn-CSV provides better forecasts than the others, whereas SS-CSV generally does well and ranks first for horizons 1--5 with respect to the density forecasts. The utility of the steady-state prior is clear from Table \ref{tab:pnfi}: while Minn-CSV and Minn-IW start out well, the performance deteriorates more rapidly with $h$ than what is manifested by the other models employing information about the steady states. We can again see that both SS-CSV and SSNG-CSV dominate Minn-CSV for all $h>0$.
\paragraph{Monthly univariate forecasting results}

Moving to the monthly variables, Table \ref{tab:cpiaucsl} presents the forecast evaluation for inflation. The results indicate that there is little to gain from using the mixed-frequency VAR for forecasting monthly inflation as compared to a monthly VAR. The relative RMSE is close to unity and few of the Diebold-Mariano tests of equal predictive ability indicate any difference between the benchmark and the mixed-frequency models.

Next, the evaluation of the forecasts of the federal funds rate is displayed in Table \ref{tab:fedfunds}. In contrast to the results for inflation, we here find large benefits from using the mixed-frequency models for forecasting the monthly federal funds rate. All three models with stochastic volatility do well with respect to both density and point forecasts, but the steady-state models have a small edge across most horizons. 

The final series we evaluate univariate forecasts for is the unemployment rate. The results are presented in Table \ref{tab:unrate}. The table reveals that mixed-frequency models are useful also for forecasting unemployment. SS-IW appears to be the better forecaster in terms of point forecasts, whereas SS-CSV provides more accurate density forecasts for all horizons. Thus, adding stochastic volatility does not improve point forecasts of the unemployment rate, but the density forecasts exhibit a substantial enhancement.

\paragraph{Forecasts evaluated against the second vintage}
To ensure that our results are not primarily driven by our choice of data to evaluate the forecasts against, Appendix \ref{app:second} presents the same tables as shown in the main text but with the evaluations carried out with the second available vintages. Qualitatively, the results remain. For the forecasts of GDP, the gains obtained by using mixed-frequency data are larger when the forecasts are evaluated against the second vintage. Occasional changes in rankings among the models occur across variables, but for the most part the rankings remain unaltered and the conclusions made so far are intact irrespective of the choice of evaluation vintage.

\section{Conclusion}
\label{sec:con}
We present a vector autoregressive model that is a synthesis of recent important contributions. Our model incorporates three main features. First, the model allows for mixed-frequency data by use of a state-space formulation. We deal with the particular mixed-frequency case involving monthly and quarterly data and solve the frequency mismatch problem by postulating a monthly VAR with missing values similar to the work by \cite{Schorfheide2015}. Second, we include prior beliefs about the steady states, or unconditional means, of the variables in the model by means of the steady-state prior developed by \cite{Villani2009}. We also employ the hierarchical formulation of the prior proposed by \cite{Louzis2019}, whose advantage is that it is only necessary to specify prior means of the steady-state parameters while the prior variances are, in turn, equipped with hyperpriors. Third, to allow for an error covariance matrix that varies over time we include as the final component the common stochastic volatility model presented by \cite{Carriero2016}. 

We estimate our model and competing alternatives using US data including ten monthly and three quarterly variables. The results show that the forecasts are generally improved by adding the three components to the benchmark VAR model. Using mixed instead of single frequencies of the data generally does not produce worse forecasts, and instead usually performs better. Including prior information about the steady states typically outperforms the corresponding alternatives that lack this information. The hierarchical steady-state prior is appealing as it allows for shrinkage to the prior means of the steady states, and is generally on par with or better than the standard steady-state prior. Finally, we find that common stochastic volatility mostly improves the accuracy of the forecasts as the models including heteroskedasticity generally outperform the models with constant volatility.

\addcontentsline{toc}{section}{References}
\bibliography{library}
\bibliographystyle{DeGruyter}

\clearpage
\appendix

\section{Posterior Moments}
\label{sec:posterior}

\paragraph{Regression and Covariance Parameters}
The moments of the posterior distributions for the regression and covariance parameters are:
\begin{align}
\overline{S}&=\underline{S}+S+(\underline{\Pi}-\hat{\Pi})\left[\underline{\Omega}_\Pi+\left(\sum_{t=1}^T\bar{Z}_{t-1}\bar{Z}_{t-1}'\right)^{-1}\right]^{-1}(\underline{\Pi}-\hat{\Pi})'\\
\hat{\Pi}&=\sum_{t=1}^T \bar{z}_t\bar{Z}_{t-1}'\left(\sum_{t=1}^T \bar{Z}_t\bar{Z}_{t-1}'\right)^{-1}\\
 S &= \sum_{t=1}^T \left(\bar{z}_t-\hat{\Pi}\bar{Z}_{t-1}\right) \left(\bar{z}_t-\hat{\Pi}\bar{Z}_{t-1}\right)'\\
\bar{\Omega}_\Pi^{-1}&=\underline{\Omega}_\Pi^{-1}+\sum_{t=1}^T\bar{Z}_{t-1}\bar{Z}_{t-1}' \\
\bar{\Pi}& = \bar{\Omega}_\Pi\left(\underline{\Pi}\,\underline{\Omega}_\Pi+\sum_{t=1}^T \bar{z}_t\bar{Z}_{t-1}'\right),
\end{align}
where $\bar{z}_t=(z_t-\Psi d_t)/\sqrt{f}_t$ and $\bar{Z}_t=\begin{pmatrix}\bar{z}_{t-1}' & \cdots & \bar{z}_{t-p}'\end{pmatrix}'$.

\paragraph{Latent Volatility}
Let $h_t=\log(f_t)$. The conditional posterior distribution of $\phi$ is
\begin{align}
\phi|h, \sigma^2&\sim \mathrm{N}(\overline{\mu}_{\phi}, \overline{\Omega}_\phi; |\phi|< 1)\\
\overline{\mu}_\phi &= \overline{\Omega}_\phi\left(\frac{\sum_{t=1}^Th_{t-1}h_t}{\sigma^2}+\frac{\underline{\mu}_\phi}{\underline{\Omega}_\phi}\right)\\
\overline{\Omega}_\phi^{-1}&=\left(\underline{\Omega}_\phi^{-1}+\frac{\sum_{t=1}^Th_{t-1}^2}{\sigma^2}\right).
\end{align}
The conditional posterior distribution of $\sigma^2$ is
\begin{align}
\sigma^2|h, \phi &\sim \mathrm{IG}(\overline{d}, \overline{\sigma}^2)\\
\overline{d}&=\underline{d}+T\\
\overline{\sigma}^2&=\sum_{t=1}^T(h_t-\phi h_{t-1})^2+\underline{d}\cdot \underline{\sigma}^2.
\end{align}

\section{Data Sources}
\label{app:data}
The IDs of the series used and their sources are shown in Table \ref{tab:sources}.
\begin{table}
\footnotesize
\centering
\begin{threeparttable}
\caption{Source and ID of Series Used}
\label{tab:sources}
\begin{tabular}{lll}
\toprule
Series  & Source  & ID \\
\midrule
Nonfarm payrolls & ALFRED & PAYEMS\\
Hours  & FRED/ALFRED & CEU0500000034\\
Unemployment rate & ALFRED  & UNRATE\\
Federal funds rate  & ALFRED& FEDFUNDS\\\vspace{0.5em}
Bond spread  & ALFRED& T10YFF\\
Stock market index  & FRED-MD& S\&P500\\
Personal consumption& ALFRED & PCE \\
Industrial production & ALFRED& INDPRO \\
Capacity utilization & ALFRED& TCU \\\vspace{0.5em}
CPI inflation & ALFRED& CPIAUCSL \\
Nonresidential inv. &ALFRED& PNFI \\
Residential inv. & ALFRED&PRFI \\
GDP growth & ALFRED& GDPC1 \\
\bottomrule
\end{tabular}
\end{threeparttable}
\end{table}

\section{Main Forecast Evaluation Tables}

\begin{table}[t]
\centering
\fontsize{8}{10}\selectfont
\begin{threeparttable}
\caption{\label{tab:lpds}Relative Joint Log Predictive Density Scores}
\begin{tabular}{llllllllll}
\toprule
Model & $h = 0$ & $h = 1$ & $h = 2$ & $h = 3$ & $h = 4$ & $h = 5$ & $h = 6$ & $h = 7$ & $h = 8$\\
\midrule
\addlinespace[0.3em]
\multicolumn{10}{l}{\textbf{Relative Joint LPDS, Quarterly}}\\
\hspace{1em}Minn-IW & -0.11 & -0.33$^{*}$ & 0.01 & 0.24 & 0.33 & 0.52 & 0.52 & 0.45 & 0.41\\
\hspace{1em}SS-IW & -0.17 & -0.48$^{*}$ & -0.22 & -0.09 & -0.08 & 0.05 & 0.02 & -0.07 & -0.13\\
\hspace{1em}SSNG-IW & -0.14 & -0.42$^{*}$ & -0.15 & -0.02 & -0.04 & 0.07 & 0.02 & -0.09 & -0.12\\
\hspace{1em}Minn-CSV & -0.36 & -1.01$^{**}$ & -0.76$^{*}$ & -0.57$^{*}$ & -0.49 & -0.22 & -0.05 & -0.06 & -0.05\\
\hspace{1em}SS-CSV & -0.42 & \textbf{-1.07}$^{**}$ & \textbf{-0.89}$^{*}$ & \textbf{-0.77}$^{*}$ & \textbf{-0.74}$^{*}$ & \textbf{-0.52} & \textbf{-0.39} & \textbf{-0.39} & \textbf{-0.39}\\
\hspace{1em}SSNG-CSV & \textbf{-0.43} & -1.07$^{**}$ & -0.86$^{*}$ & -0.73$^{*}$ & -0.69$^{*}$ & -0.44 & -0.32 & -0.36 & -0.36\\
\addlinespace[0.3em]
\multicolumn{10}{l}{\textbf{Relative Joint LPDS, Monthly}}\\
\hspace{1em}Minn-IW &  & -1.74$^{**}$ & -1.49$^{**}$ & -1.21$^{**}$ & -1.04$^{*}$ & -1.14$^{**}$ & -1.03$^{**}$ & -1.02$^{**}$ & -0.88$^{**}$\\
\hspace{1em}SS-IW &  & -1.85$^{**}$ & -1.44$^{**}$ & -1.12$^{**}$ & -1.04$^{**}$ & -1.14$^{**}$ & -1.04$^{**}$ & -1.05$^{**}$ & -0.95$^{**}$\\
\hspace{1em}SSNG-IW &  & -1.83$^{**}$ & -1.47$^{**}$ & -1.19$^{**}$ & -1.03$^{*}$ & -1.22$^{**}$ & -1.12$^{**}$ & -1.08$^{**}$ & -0.95$^{**}$\\
\hspace{1em}Minn-CSV &  & -1.96$^{*}$ & -2.93$^{*}$ & -3.01$^{*}$ & -3.03$^{*}$ & -2.98$^{*}$ & -2.77$^{*}$ & -2.62$^{*}$ & -2.29$^{*}$\\
\hspace{1em}SS-CSV &  & \textbf{-2.17}$^{*}$ & -3.01$^{*}$ & -3.00$^{*}$ & -3.07$^{*}$ & \textbf{-3.13}$^{*}$ & \textbf{-3.01}$^{*}$ & \textbf{-2.98}$^{*}$ & \textbf{-2.65}$^{*}$\\
\hspace{1em}SSNG-CSV &  & -2.07$^{*}$ & \textbf{-3.07}$^{*}$ & \textbf{-3.03}$^{*}$ & \textbf{-3.09}$^{*}$ & -3.13$^{*}$ & -2.97$^{*}$ & -2.86$^{*}$ & -2.53$^{*}$\\
\bottomrule
\end{tabular}
\begin{tablenotes}
\item \textit{Note: } 
\item The forecast horizons $h$ refer to quarters and months, respectively, for the two sets of variables. The scores in the table display the score of the model in the first column minus the score of the benchmark model, whereby negative entries indicate that the mixed-frequency model is superior. Bold entires show the minimum in each column. The benchmark model for the quarterly set of variables is a VAR(4) including all 13 variables aggregated to the quarterly frequency. For the monthly LPDS, the benchmark model is a VAR(12) including the the ten monthly variables. For both cases, the steady-state prior with a constant error covariance matrix is used. Two stars ($^{**}$) indicate that the Diebold-Mariano test of equal predictive ability is significant at the 1 percent level, whereas a single star indicates significance at the 10 percent level. The test employs the modifications proposed by \cite{Harvey1997}.
\end{tablenotes}
\end{threeparttable}
\end{table}

\begin{table}[t]
\centering
\fontsize{8}{10}\selectfont
\begin{threeparttable}
\caption{\label{tab:gdpc1}GDP: Forecast Evaluation}
\begin{tabular}{llllllllll}
\toprule
Model & $h = 0$ & $h = 1$ & $h = 2$ & $h = 3$ & $h = 4$ & $h = 5$ & $h = 6$ & $h = 7$ & $h = 8$\\
\midrule
\addlinespace[0.3em]
\multicolumn{10}{l}{\textbf{Relative LPDS} (model in 1st column $-$ benchmark)}\\
\hspace{1em}Minn-IW & -0.23$^{*}$ & -0.09 & 0.11 & 0.18 & 0.14 & 0.18 & 0.14 & 0.11 & 0.08\\
\hspace{1em}SS-IW & -0.23$^{*}$ & -0.13 & 0.05 & \textbf{0.06} & \textbf{0.00} & \textbf{0.01} & \textbf{-0.02} & \textbf{-0.05} & \textbf{-0.10}$^{*}$\\
\hspace{1em}SSNG-IW & -0.23$^{*}$ & -0.10 & 0.10 & 0.15 & 0.10 & 0.11 & 0.04 & 0.01 & -0.04\\
\hspace{1em}Minn-CSV & -0.27 & -0.24$^{*}$ & 0.01 & 0.15 & 0.16 & 0.20 & 0.19 & 0.16 & 0.09\\
\hspace{1em}SS-CSV & -0.27 & \textbf{-0.26}$^{*}$ & \textbf{-0.02} & 0.10 & 0.08 & 0.12 & 0.10 & 0.06 & 0.00\\
\hspace{1em}SSNG-CSV & \textbf{-0.27} & -0.25$^{*}$ & 0.00 & 0.13 & 0.14 & 0.17 & 0.14 & 0.10 & 0.03\\
\addlinespace[0.3em]
\multicolumn{10}{l}{\textbf{Relative RMSE} (model in 1st column $/$ benchmark)}\\
\hspace{1em}Minn-IW & 0.90$^{**}$ & 0.95$^{*}$ & 1.05 & 1.11 & 1.07 & 1.10 & 1.07 & 1.06 & 1.04\\
\hspace{1em}SS-IW & \textbf{0.90}$^{**}$ & \textbf{0.94}$^{*}$ & 1.03 & \textbf{1.05} & \textbf{1.00} & \textbf{1.01} & \textbf{0.98} & \textbf{0.96}$^{*}$ & \textbf{0.95}$^{*}$\\
\hspace{1em}SSNG-IW & 0.90$^{**}$ & 0.94$^{*}$ & 1.05 & 1.10 & 1.05 & 1.07 & 1.03 & 1.01 & 0.99\\
\hspace{1em}Minn-CSV & 0.92 & 0.95 & 1.03 & 1.13 & 1.10 & 1.12 & 1.08 & 1.06 & 1.03\\
\hspace{1em}SS-CSV & 0.92 & 0.94 & \textbf{1.01} & 1.08 & 1.02 & 1.04 & 1.01 & 0.97 & 0.95\\
\hspace{1em}SSNG-CSV & 0.92 & 0.94 & 1.02 & 1.12 & 1.08 & 1.09 & 1.05 & 1.01 & 0.99\\
\bottomrule
\end{tabular}
\begin{tablenotes}
\item \textit{Note: } 
\item The forecast horizon $h$ denotes quarters. Negative LPDS entries indicate that the mixed-frequency model is superior in terms of density forecasting and values of the RMSE below 1 indicate better point forecasts. Bold entries show the minimum per column. The benchmark model is a VAR(4) including all 13 variables aggregated to the quarterly frequency using the steady-state prior with a constant error covariance matrix. Two stars ($^{**}$) indicate that the Diebold-Mariano test of equal predictive ability is significant at the 1 percent level, whereas a single star indicates significance at the 10 percent level. The test employs the modifications proposed by \cite{Harvey1997}.
\end{tablenotes}
\end{threeparttable}
\end{table}

\begin{table}[t]
\centering
\fontsize{8}{10}\selectfont
\begin{threeparttable}

\caption{\label{tab:prfi}Residential Investment: Forecast Evaluation}
\begin{tabular}{llllllllll}
\toprule
Model & $h = 0$ & $h = 1$ & $h = 2$ & $h = 3$ & $h = 4$ & $h = 5$ & $h = 6$ & $h = 7$ & $h = 8$\\
\midrule
\addlinespace[0.3em]
\multicolumn{10}{l}{\textbf{Relative LPDS} (model in 1st column $-$ benchmark)}\\
\hspace{1em}Minn-IW & 0.07 & 0.00 & 0.22 & 0.16 & 0.14 & 0.12 & 0.09 & 0.14 & 0.18\\
\hspace{1em}SS-IW & -0.01 & -0.08 & 0.09 & 0.02 & -0.01 & -0.03 & -0.08 & -0.04 & -0.01\\
\hspace{1em}SSNG-IW & 0.03 & -0.03 & 0.12 & 0.00 & -0.09 & -0.18 & -0.25$^{*}$ & -0.22$^{*}$ & -0.19$^{*}$\\
\hspace{1em}Minn-CSV & -0.10 & -0.49$^{*}$ & -0.38$^{*}$ & -0.42$^{*}$ & -0.44$^{*}$ & -0.38$^{*}$ & -0.36$^{*}$ & -0.32$^{*}$ & -0.28$^{*}$\\
\hspace{1em}SS-CSV & \textbf{-0.17} & \textbf{-0.54}$^{*}$ & \textbf{-0.46}$^{*}$ & \textbf{-0.53}$^{*}$ & \textbf{-0.56}$^{**}$ & -0.53$^{*}$ & -0.53$^{*}$ & -0.48$^{*}$ & -0.46$^{*}$\\
\hspace{1em}SSNG-CSV & -0.17 & -0.54$^{*}$ & -0.43$^{*}$ & -0.51$^{*}$ & -0.55$^{**}$ & \textbf{-0.53}$^{*}$ & \textbf{-0.56}$^{*}$ & \textbf{-0.52}$^{*}$ & \textbf{-0.51}$^{*}$\\
\addlinespace[0.3em]
\multicolumn{10}{l}{\textbf{Relative RMSE} (model in 1st column $/$ benchmark)}\\
\hspace{1em}Minn-IW & 0.92$^{**}$ & 0.96$^{**}$ & 1.03 & 1.02 & 1.01 & 1.03 & 1.03 & 1.06 & 1.06\\
\hspace{1em}SS-IW & 0.90$^{**}$ & 0.92$^{**}$ & 0.99 & 0.97 & 0.97 & 0.98 & 0.99$^{*}$ & 1.01 & 1.01\\
\hspace{1em}SSNG-IW & 0.92$^{**}$ & 0.94$^{**}$ & 1.00 & 0.98 & 0.96 & 0.97 & 0.96 & 0.98 & 0.98\\
\hspace{1em}Minn-CSV & 0.88$^{**}$ & 0.90$^{**}$ & 0.96 & 0.94 & 0.94$^{*}$ & 0.96 & 0.95$^{*}$ & 0.98 & 1.00\\
\hspace{1em}SS-CSV & \textbf{0.87}$^{**}$ & \textbf{0.90}$^{**}$ & \textbf{0.95} & \textbf{0.91}$^{*}$ & \textbf{0.92}$^{*}$ & \textbf{0.93} & 0.92$^{*}$ & 0.95 & \textbf{0.96}\\
\hspace{1em}SSNG-CSV & 0.88$^{**}$ & 0.90$^{**}$ & 0.95 & 0.93 & 0.92$^{*}$ & 0.94 & \textbf{0.92}$^{*}$ & \textbf{0.95} & 0.96\\
\bottomrule
\end{tabular}
\begin{tablenotes}
\item \textit{Note: } 
\item The forecast horizon $h$ denotes quarters. Negative LPDS entries indicate that the mixed-frequency model is superior in terms of density forecasting and values of the RMSE below 1 indicate better point forecasts. Bold entries show the minimum per column. The benchmark model is a VAR(4) including all 13 variables aggregated to the quarterly frequency using the steady-state prior with a constant error covariance matrix. Two stars ($^{**}$) indicate that the Diebold-Mariano test of equal predictive ability is significant at the 1 percent level, whereas a single star indicates significance at the 10 percent level. The test employs the modifications proposed by \cite{Harvey1997}.
\end{tablenotes}
\end{threeparttable}
\end{table}

\begin{table}[t]
\centering
\fontsize{8}{10}\selectfont
\begin{threeparttable}

\caption{\label{tab:pnfi}Non-Residential Investment: Forecast Evaluation}
\begin{tabular}{llllllllll}
\toprule
Model & $h = 0$ & $h = 1$ & $h = 2$ & $h = 3$ & $h = 4$ & $h = 5$ & $h = 6$ & $h = 7$ & $h = 8$\\
\midrule
\addlinespace[0.3em]
\multicolumn{10}{l}{\textbf{Relative LPDS} (model in 1st column $-$ benchmark)}\\
\hspace{1em}Minn-IW & -0.09$^{*}$ & -0.38$^{**}$ & -0.16$^{*}$ & 0.01 & 0.13 & 0.24 & 0.28 & 0.21 & 0.16\\
\hspace{1em}SS-IW & -0.10$^{*}$ & -0.42$^{**}$ & -0.22$^{*}$ & -0.09 & -0.02 & 0.08 & \textbf{0.11} & \textbf{0.03} & \textbf{-0.03}\\
\hspace{1em}SSNG-IW & -0.09$^{*}$ & -0.41$^{**}$ & -0.20$^{*}$ & -0.05 & 0.05 & 0.16 & 0.20 & 0.11 & 0.05\\
\hspace{1em}Minn-CSV & \textbf{-0.12} & -0.45$^{**}$ & -0.24 & -0.06 & 0.06 & 0.21 & 0.32 & 0.31 & 0.26\\
\hspace{1em}SS-CSV & -0.11 & \textbf{-0.47}$^{**}$ & \textbf{-0.32}$^{*}$ & \textbf{-0.17} & \textbf{-0.06} & \textbf{0.06} & 0.16 & 0.13 & 0.09\\
\hspace{1em}SSNG-CSV & -0.12$^{*}$ & -0.46$^{**}$ & -0.29$^{*}$ & -0.14 & -0.02 & 0.13 & 0.25 & 0.22 & 0.18\\
\addlinespace[0.3em]
\multicolumn{10}{l}{\textbf{Relative RMSE} (model in 1st column $/$ benchmark)}\\
\hspace{1em}Minn-IW & 0.97 & 0.83$^{**}$ & 0.93 & 0.98 & 1.02 & 1.09 & 1.12 & 1.11 & 1.08\\
\hspace{1em}SS-IW & 0.95 & 0.82$^{**}$ & 0.92$^{*}$ & 0.95 & \textbf{0.98} & \textbf{1.03} & \textbf{1.04} & \textbf{1.00} & \textbf{0.98}$^{*}$\\
\hspace{1em}SSNG-IW & 0.97 & 0.83$^{**}$ & 0.91$^{*}$ & 0.95 & 0.99 & 1.05 & 1.08 & 1.06 & 1.03\\
\hspace{1em}Minn-CSV & \textbf{0.93} & 0.83$^{**}$ & 0.93 & 1.01 & 1.07 & 1.15 & 1.23 & 1.22 & 1.19\\
\hspace{1em}SS-CSV & 0.94 & \textbf{0.81}$^{**}$ & \textbf{0.88}$^{*}$ & \textbf{0.94} & 0.99 & 1.04 & 1.08 & 1.06 & 1.03\\
\hspace{1em}SSNG-CSV & 0.93 & 0.81$^{**}$ & 0.90 & 0.97 & 1.03 & 1.10 & 1.17 & 1.16 & 1.13\\
\bottomrule
\end{tabular}
\begin{tablenotes}
\item \textit{Note: } 
\item The forecast horizon $h$ denotes quarters. Negative LPDS entries indicate that the mixed-frequency model is superior in terms of density forecasting and values of the RMSE below 1 indicate better point forecasts. Bold entries show the minimum per column. The benchmark model is a VAR(4) including all 13 variables aggregated to the quarterly frequency using the steady-state prior with a constant error covariance matrix. Two stars ($^{**}$) indicate that the Diebold-Mariano test of equal predictive ability is significant at the 1 percent level, whereas a single star indicates significance at the 10 percent level. The test employs the modifications proposed by \cite{Harvey1997}.
\end{tablenotes}
\end{threeparttable}
\end{table}

\begin{table}[t]
\centering
\fontsize{8}{10}\selectfont
\begin{threeparttable}

\caption{\label{tab:cpiaucsl}Inflation: Forecast Evaluation}
\begin{tabular}{llllllllll}
\toprule
Model & $h = 0$ & $h = 1$ & $h = 2$ & $h = 3$ & $h = 4$ & $h = 5$ & $h = 6$ & $h = 7$ & $h = 8$\\
\midrule
\addlinespace[0.3em]
\multicolumn{10}{l}{\textbf{Relative LPDS} (model in 1st column $-$ benchmark)}\\
\hspace{1em}Minn-IW & 0.07 & -0.09 & -0.02 & -0.01 & 0.01 & -0.02 & -0.01 & -0.01 & -0.02\\
\hspace{1em}SS-IW & 0.06 & -0.10 & 0.00 & 0.00 & 0.03 & -0.00 & -0.00 & 0.00 & -0.01\\
\hspace{1em}SSNG-IW & 0.05 & -0.10 & -0.01 & -0.01 & 0.04 & -0.01 & -0.04 & -0.02 & -0.03$^{*}$\\
\hspace{1em}Minn-CSV & -0.22 & -0.42 & -0.31 & \textbf{-0.33} & \textbf{-0.34} & -0.30 & -0.29 & -0.32 & -0.30\\
\hspace{1em}SS-CSV & -0.22 & \textbf{-0.43} & \textbf{-0.32} & -0.31 & -0.33 & \textbf{-0.32} & -0.31 & \textbf{-0.36} & -0.32\\
\hspace{1em}SSNG-CSV & \textbf{-0.22} & -0.43 & -0.31 & -0.32 & -0.33 & -0.29 & \textbf{-0.33} & -0.36 & \textbf{-0.34}\\
\addlinespace[0.3em]
\multicolumn{10}{l}{\textbf{Relative RMSE} (model in 1st column $/$ benchmark)}\\
\hspace{1em}Minn-IW & 1.02 & \textbf{0.98} & 0.99 & 0.99 & 1.00 & 1.00 & 1.00 & 1.00 & 1.00\\
\hspace{1em}SS-IW & \textbf{1.02} & 0.98 & 1.00 & 1.00 & 1.01 & 1.01 & 1.00 & 1.00 & 1.00\\
\hspace{1em}SSNG-IW & 1.02 & 0.98 & 0.99 & 0.99 & 1.00 & 1.00 & 1.00 & 1.00 & 1.00\\
\hspace{1em}Minn-CSV & 1.05 & 0.99 & 0.97 & 0.96 & \textbf{0.97} & \textbf{0.97} & 0.97$^{*}$ & 0.98$^{*}$ & 0.99\\
\hspace{1em}SS-CSV & 1.04 & 0.98 & 0.97 & 0.97 & 0.97$^{*}$ & 0.97 & 0.97$^{*}$ & \textbf{0.97}$^{*}$ & \textbf{0.98}\\
\hspace{1em}SSNG-CSV & 1.04 & 0.99 & \textbf{0.97} & \textbf{0.96} & 0.97 & 0.97 & \textbf{0.97}$^{*}$ & 0.98$^{*}$ & 0.98\\
\bottomrule
\end{tabular}
\begin{tablenotes}
\item \textit{Note: } 
\item The forecast horizon $h$ denotes months. Negative LPDS entries indicate that the mixed-frequency model is superior in terms of density forecasting and values of the RMSE below 1 indicate better point forecasts. Bold entries show the minimum per column. The benchmark model is a VAR(12) including the ten monthly variables using the steady-state prior with a constant error covariance matrix. Two stars ($^{**}$) indicate that the Diebold-Mariano test of equal predictive ability is significant at the 1 percent level, whereas a single star indicates significance at the 10 percent level. The test employs the modifications proposed by \cite{Harvey1997}.
\end{tablenotes}
\end{threeparttable}
\end{table}

\begin{table}[t]
\centering
\fontsize{8}{10}\selectfont
\begin{threeparttable}

\caption{\label{tab:fedfunds}Federal Funds Rate: Forecast Evaluation}
\begin{tabular}{lllllllll}
\toprule
Model & $h = 1$ & $h = 2$ & $h = 3$ & $h = 4$ & $h = 5$ & $h = 6$ & $h = 7$ & $h = 8$\\
\midrule
\addlinespace[0.3em]
\multicolumn{9}{l}{\textbf{Relative LPDS} (model in 1st column $-$ benchmark)}\\
\hspace{1em}Minn-IW & -0.91$^{**}$ & -0.50$^{**}$ & -0.32$^{*}$ & -0.24$^{*}$ & -0.20 & -0.18 & -0.17 & -0.17\\
\hspace{1em}SS-IW & -0.93$^{**}$ & -0.53$^{**}$ & -0.35$^{**}$ & -0.27$^{**}$ & -0.23$^{*}$ & -0.21$^{*}$ & -0.21$^{*}$ & -0.21$^{*}$\\
\hspace{1em}SSNG-IW & -0.92$^{**}$ & -0.52$^{**}$ & -0.35$^{**}$ & -0.27$^{**}$ & -0.23$^{**}$ & -0.21$^{*}$ & -0.20$^{*}$ & -0.20$^{*}$\\
\hspace{1em}Minn-CSV & -1.45$^{**}$ & -1.04$^{**}$ & -0.80$^{**}$ & -0.63$^{**}$ & -0.52$^{**}$ & -0.44$^{**}$ & -0.38$^{**}$ & -0.34$^{*}$\\
\hspace{1em}SS-CSV & \textbf{-1.47}$^{**}$ & \textbf{-1.06}$^{**}$ & \textbf{-0.82}$^{**}$ & \textbf{-0.64}$^{**}$ & \textbf{-0.53}$^{**}$ & \textbf{-0.45}$^{**}$ & \textbf{-0.39}$^{**}$ & \textbf{-0.35}$^{**}$\\
\hspace{1em}SSNG-CSV & -1.46$^{**}$ & -1.05$^{**}$ & -0.81$^{**}$ & -0.64$^{**}$ & -0.52$^{**}$ & -0.44$^{**}$ & -0.37$^{**}$ & -0.34$^{**}$\\
\addlinespace[0.3em]
\multicolumn{9}{l}{\textbf{Relative RMSE} (model in 1st column $/$ benchmark)}\\
\hspace{1em}Minn-IW & 0.58$^{**}$ & 0.75$^{*}$ & 0.85 & 0.90 & 0.93 & 0.94 & 0.94 & 0.94\\
\hspace{1em}SS-IW & 0.56$^{**}$ & 0.72$^{*}$ & 0.81$^{*}$ & 0.86$^{*}$ & 0.89$^{*}$ & 0.90$^{*}$ & 0.90$^{*}$ & 0.90$^{*}$\\
\hspace{1em}SSNG-IW & 0.57$^{**}$ & 0.72$^{*}$ & 0.81$^{*}$ & 0.86$^{*}$ & 0.89$^{*}$ & 0.90$^{*}$ & 0.91$^{*}$ & 0.90$^{**}$\\
\hspace{1em}Minn-CSV & 0.53$^{**}$ & 0.68$^{*}$ & 0.76$^{*}$ & 0.81 & 0.84 & 0.86 & 0.87 & 0.87\\
\hspace{1em}SS-CSV & \textbf{0.51}$^{**}$ & \textbf{0.65}$^{*}$ & \textbf{0.73}$^{*}$ & \textbf{0.78}$^{*}$ & \textbf{0.82}$^{*}$ & \textbf{0.83}$^{*}$ & \textbf{0.85}$^{*}$ & \textbf{0.85}\\
\hspace{1em}SSNG-CSV & 0.52$^{**}$ & 0.67$^{*}$ & 0.75$^{*}$ & 0.80$^{*}$ & 0.83$^{*}$ & 0.84$^{*}$ & 0.85 & 0.86\\
\bottomrule
\end{tabular}
\begin{tablenotes}
\item \textit{Note: } 
\item The forecast horizon $h$ denotes months. Negative LPDS entries indicate that the mixed-frequency model is superior in terms of density forecasting and values of the RMSE below 1 indicate better point forecasts. Bold entries show the minimum per column. The benchmark model is a VAR(12) including the ten monthly variables using the steady-state prior with a constant error covariance matrix. Two stars ($^{**}$) indicate that the Diebold-Mariano test of equal predictive ability is significant at the 1 percent level, whereas a single star indicates significance at the 10 percent level. The test employs the modifications proposed by \cite{Harvey1997}.
\end{tablenotes}
\end{threeparttable}
\end{table}

\begin{table}[t]
\centering
\fontsize{8}{10}\selectfont
\begin{threeparttable}

\caption{\label{tab:unrate}Unemployment: Forecast Evaluation}
\begin{tabular}{lllllllll}
\toprule
Model & $h = 1$ & $h = 2$ & $h = 3$ & $h = 4$ & $h = 5$ & $h = 6$ & $h = 7$ & $h = 8$\\
\midrule
\addlinespace[0.3em]
\multicolumn{9}{l}{\textbf{Relative LPDS} (model in 1st column $-$ benchmark)}\\
\hspace{1em}Minn-IW & -0.43 & -0.34 & -0.25 & -0.28 & -0.34 & -0.30 & -0.32$^{*}$ & -0.29\\
\hspace{1em}SS-IW & -0.46 & -0.39 & -0.31 & -0.35 & -0.41$^{*}$ & -0.38 & -0.40 & -0.37\\
\hspace{1em}SSNG-IW & -0.43 & -0.33 & -0.21 & -0.24 & -0.29 & -0.25 & -0.26 & -0.22\\
\hspace{1em}Minn-CSV & -0.48 & -0.48 & -0.51 & -0.61 & -0.74 & -0.77 & -0.82 & -0.84\\
\hspace{1em}SS-CSV & \textbf{-0.49} & \textbf{-0.50} & \textbf{-0.52} & \textbf{-0.63} & \textbf{-0.77} & \textbf{-0.80} & \textbf{-0.87} & \textbf{-0.89}\\
\hspace{1em}SSNG-CSV & -0.47 & -0.47 & -0.49 & -0.59 & -0.72 & -0.74 & -0.80 & -0.82\\
\addlinespace[0.3em]
\multicolumn{9}{l}{\textbf{Relative RMSE} (model in 1st column $/$ benchmark)}\\
\hspace{1em}Minn-IW & 0.78$^{**}$ & 0.83$^{*}$ & 0.87$^{*}$ & 0.87 & 0.86 & 0.88 & 0.89 & 0.90\\
\hspace{1em}SS-IW & \textbf{0.77}$^{**}$ & \textbf{0.81}$^{*}$ & \textbf{0.84} & \textbf{0.84} & \textbf{0.84} & \textbf{0.86} & \textbf{0.87} & \textbf{0.88}\\
\hspace{1em}SSNG-IW & 0.79$^{**}$ & 0.84$^{*}$ & 0.87$^{*}$ & 0.87 & 0.87 & 0.89 & 0.89 & 0.91\\
\hspace{1em}Minn-CSV & 0.82$^{**}$ & 0.87$^{*}$ & 0.90 & 0.90 & 0.89 & 0.89 & 0.90 & 0.91\\
\hspace{1em}SS-CSV & 0.81$^{**}$ & 0.86$^{*}$ & 0.89 & 0.89 & 0.88 & 0.89 & 0.89 & 0.90\\
\hspace{1em}SSNG-CSV & 0.83$^{*}$ & 0.88$^{*}$ & 0.91 & 0.91 & 0.90 & 0.91 & 0.91 & 0.92\\
\bottomrule
\end{tabular}
\begin{tablenotes}
\item \textit{Note: } 
\item The forecast horizon $h$ denotes months. Negative LPDS entries indicate that the mixed-frequency model is superior in terms of density forecasting and values of the RMSE below 1 indicate better point forecasts. Bold entries show the minimum per column. The benchmark model is a VAR(12) including the ten monthly variables using the steady-state prior with a constant error covariance matrix. Two stars ($^{**}$) indicate that the Diebold-Mariano test of equal predictive ability is significant at the 1 percent level, whereas a single star indicates significance at the 10 percent level. The test employs the modifications proposed by \cite{Harvey1997}.
\end{tablenotes}
\end{threeparttable}
\end{table}

\clearpage

\section{Forecast Evaluation Tables (Second Vintage)}
\label{app:second}
The tables in the main text present the results of the forecast evaluation when evaluated with respect to the most recent vintage. The tables in this Appendix (Table \ref{tab:gdpc12}--\ref{tab:unrate2}) present the same evaluations but conducted with respect to the second available vintage. Because the federal funds rate is not revised, the second vintage is the same as the most recent vintage. Therefore, the results when evaluating the forecasts against the second vintage are identical to Table \ref{tab:fedfunds} and therefore not reproduced again here.

\begin{table}[t]
\centering
\fontsize{8}{10}\selectfont
\begin{threeparttable}

\caption{\label{tab:gdpc12}GDP: Forecast Evaluation (Second Vintage)}
\begin{tabular}{llllllllll}
\toprule
Model & $h = 0$ & $h = 1$ & $h = 2$ & $h = 3$ & $h = 4$ & $h = 5$ & $h = 6$ & $h = 7$ & $h = 8$\\
\midrule
\addlinespace[0.3em]
\multicolumn{10}{l}{\textbf{Relative LPDS} (model in 1st column $-$ benchmark)}\\
\hspace{1em}Minn-IW & \textbf{-0.34}$^{**}$ & -0.15 & 0.03 & 0.12 & 0.14 & 0.13 & 0.09 & 0.09 & 0.09\\
\hspace{1em}SS-IW & -0.31$^{*}$ & -0.16$^{*}$ & 0.01 & \textbf{0.04} & \textbf{0.04} & \textbf{-0.00} & \textbf{-0.05}$^{*}$ & \textbf{-0.05}$^{**}$ & \textbf{-0.05}\\
\hspace{1em}SSNG-IW & -0.33$^{**}$ & -0.15 & 0.03 & 0.11 & 0.12 & 0.08 & 0.02 & 0.01 & 0.00\\
\hspace{1em}Minn-CSV & -0.28 & -0.27$^{**}$ & \textbf{-0.06} & 0.10 & 0.17 & 0.19 & 0.17 & 0.17 & 0.16\\
\hspace{1em}SS-CSV & -0.24 & \textbf{-0.28}$^{**}$ & -0.05 & 0.08 & 0.14 & 0.14 & 0.10 & 0.10 & 0.10\\
\hspace{1em}SSNG-CSV & -0.26 & -0.27$^{**}$ & -0.05 & 0.10 & 0.17 & 0.19 & 0.15 & 0.14 & 0.11\\
\addlinespace[0.3em]
\multicolumn{10}{l}{\textbf{Relative RMSE} (model in 1st column $/$ benchmark)}\\
\hspace{1em}Minn-IW & \textbf{0.83}$^{**}$ & 0.92$^{**}$ & 1.03 & 1.09 & 1.11 & 1.10 & 1.07 & 1.07 & 1.06\\
\hspace{1em}SS-IW & 0.84$^{**}$ & 0.92$^{**}$ & 1.03 & 1.06 & \textbf{1.04} & \textbf{1.00} & \textbf{0.97} & \textbf{0.96} & \textbf{0.96}\\
\hspace{1em}SSNG-IW & 0.83$^{**}$ & 0.92$^{**}$ & 1.03 & 1.08 & 1.10 & 1.06 & 1.03 & 1.02 & 1.01\\
\hspace{1em}Minn-CSV & 0.85$^{**}$ & 0.90$^{*}$ & 0.98 & 1.09 & 1.14 & 1.12 & 1.08 & 1.08 & 1.06\\
\hspace{1em}SS-CSV & 0.87$^{**}$ & 0.90$^{**}$ & 0.97 & \textbf{1.05} & 1.07 & 1.04 & 1.00 & 0.99 & 0.99\\
\hspace{1em}SSNG-CSV & 0.85$^{**}$ & \textbf{0.90}$^{**}$ & \textbf{0.97} & 1.08 & 1.12 & 1.10 & 1.05 & 1.04 & 1.02\\
\bottomrule
\end{tabular}
\begin{tablenotes}
\item \textit{Note: } 
\item The forecast horizon $h$ denotes quarters. Negative LPDS entries indicate that the mixed-frequency model is superior in terms of density forecasting and values of the RMSE below 1 indicate better point forecasts. Bold entries show the minimum per column. The benchmark model is a VAR(4) including all 13 variables aggregated to the quarterly frequency using the steady-state prior with a constant error covariance matrix. Two stars ($^{**}$) indicate that the Diebold-Mariano test of equal predictive ability is significant at the 1 percent level, whereas a single star indicates significance at the 10 percent level. The test employs the modifications proposed by \cite{Harvey1997}.
\end{tablenotes}
\end{threeparttable}
\end{table}

\begin{table}[t]
\centering
\fontsize{8}{10}\selectfont
\begin{threeparttable}

\caption{\label{tab:prfi2}Residential Investment: Forecast Evaluation (Second Vintage)}
\begin{tabular}{llllllllll}
\toprule
Model & $h = 0$ & $h = 1$ & $h = 2$ & $h = 3$ & $h = 4$ & $h = 5$ & $h = 6$ & $h = 7$ & $h = 8$\\
\midrule
\addlinespace[0.3em]
\multicolumn{10}{l}{\textbf{Relative LPDS} (model in 1st column $-$ benchmark)}\\
\hspace{1em}Minn-IW & 0.01 & -0.11$^{*}$ & 0.14 & 0.14 & 0.14 & 0.11 & 0.08 & 0.14 & 0.18\\
\hspace{1em}SS-IW & -0.05 & -0.18$^{*}$ & 0.03 & 0.01 & -0.01 & -0.05 & -0.09 & -0.06 & -0.01\\
\hspace{1em}SSNG-IW & -0.00 & -0.12$^{*}$ & 0.07 & 0.01 & -0.06 & -0.16 & -0.22$^{*}$ & -0.20$^{*}$ & -0.15\\
\hspace{1em}Minn-CSV & -0.18 & -0.53$^{*}$ & -0.38$^{*}$ & -0.36$^{*}$ & -0.37$^{*}$ & -0.31$^{*}$ & -0.28$^{*}$ & -0.23 & -0.17\\
\hspace{1em}SS-CSV & \textbf{-0.24} & \textbf{-0.56}$^{*}$ & \textbf{-0.44}$^{*}$ & \textbf{-0.45}$^{*}$ & \textbf{-0.46}$^{*}$ & \textbf{-0.44}$^{*}$ & -0.43$^{*}$ & -0.38$^{*}$ & -0.33\\
\hspace{1em}SSNG-CSV & -0.23 & -0.56$^{*}$ & -0.42$^{*}$ & -0.43$^{*}$ & -0.46$^{*}$ & -0.43$^{*}$ & \textbf{-0.45}$^{*}$ & \textbf{-0.41}$^{*}$ & \textbf{-0.37}\\
\addlinespace[0.3em]
\multicolumn{10}{l}{\textbf{Relative RMSE} (model in 1st column $/$ benchmark)}\\
\hspace{1em}Minn-IW & 0.93$^{*}$ & 0.96$^{*}$ & 1.03 & 1.02 & 1.02 & 1.03 & 1.04 & 1.06 & 1.06\\
\hspace{1em}SS-IW & 0.90$^{*}$ & 0.92$^{**}$ & 0.99 & 0.98 & 0.97 & 0.98 & 0.98$^{*}$ & 1.00 & 1.01\\
\hspace{1em}SSNG-IW & 0.92$^{*}$ & 0.95$^{*}$ & 1.01 & 0.98 & 0.97 & 0.97 & 0.96 & 0.98 & 0.98\\
\hspace{1em}Minn-CSV & 0.89$^{*}$ & 0.91$^{**}$ & 0.96 & 0.94 & 0.95$^{*}$ & 0.96 & 0.96 & 1.00 & 1.01\\
\hspace{1em}SS-CSV & \textbf{0.88}$^{**}$ & \textbf{0.91}$^{**}$ & \textbf{0.95} & \textbf{0.91}$^{*}$ & \textbf{0.92}$^{*}$ & \textbf{0.93} & \textbf{0.93}$^{*}$ & \textbf{0.96} & \textbf{0.97}\\
\hspace{1em}SSNG-CSV & 0.88$^{*}$ & 0.91$^{**}$ & 0.96 & 0.93 & 0.93$^{*}$ & 0.94 & 0.93 & 0.97 & 0.97\\
\bottomrule
\end{tabular}
\begin{tablenotes}
\item \textit{Note: } 
\item The forecast horizon $h$ denotes quarters. Negative LPDS entries indicate that the mixed-frequency model is superior in terms of density forecasting and values of the RMSE below 1 indicate better point forecasts. Bold entries show the minimum per column. The benchmark model is a VAR(4) including all 13 variables aggregated to the quarterly frequency using the steady-state prior with a constant error covariance matrix. Two stars ($^{**}$) indicate that the Diebold-Mariano test of equal predictive ability is significant at the 1 percent level, whereas a single star indicates significance at the 10 percent level. The test employs the modifications proposed by \cite{Harvey1997}.
\end{tablenotes}
\end{threeparttable}
\end{table}

\begin{table}[t]
\centering
\fontsize{8}{10}\selectfont
\begin{threeparttable}

\caption{\label{tab:pnfi2}Non-Residential Investment: Forecast Evaluation (Second Vintage)}
\begin{tabular}{llllllllll}
\toprule
Model & $h = 0$ & $h = 1$ & $h = 2$ & $h = 3$ & $h = 4$ & $h = 5$ & $h = 6$ & $h = 7$ & $h = 8$\\
\midrule
\addlinespace[0.3em]
\multicolumn{10}{l}{\textbf{Relative LPDS} (model in 1st column $-$ benchmark)}\\
\hspace{1em}Minn-IW & -0.07 & -0.52$^{*}$ & -0.08 & 0.05 & 0.19 & 0.34 & 0.38 & 0.29 & 0.23\\
\hspace{1em}SS-IW & -0.11$^{*}$ & -0.62$^{*}$ & -0.21$^{*}$ & -0.11 & -0.03 & 0.09 & 0.12 & \textbf{0.02} & \textbf{-0.04}\\
\hspace{1em}SSNG-IW & -0.09$^{*}$ & -0.57$^{*}$ & -0.15$^{*}$ & -0.02 & 0.10 & 0.24 & 0.27 & 0.16 & 0.10\\
\hspace{1em}Minn-CSV & -0.16$^{*}$ & -0.74$^{*}$ & -0.37 & -0.17 & -0.04 & 0.18 & 0.33 & 0.29 & 0.25\\
\hspace{1em}SS-CSV & \textbf{-0.19}$^{*}$ & \textbf{-0.81}$^{*}$ & \textbf{-0.49} & \textbf{-0.33} & \textbf{-0.22} & \textbf{-0.04} & \textbf{0.08} & 0.03 & 0.01\\
\hspace{1em}SSNG-CSV & -0.17$^{*}$ & -0.78$^{*}$ & -0.44 & -0.27 & -0.14 & 0.09 & 0.23 & 0.17 & 0.14\\
\addlinespace[0.3em]
\multicolumn{10}{l}{\textbf{Relative RMSE} (model in 1st column $/$ benchmark)}\\
\hspace{1em}Minn-IW & 0.99 & 0.84$^{*}$ & 0.96 & 0.99 & 1.05 & 1.11 & 1.14 & 1.12 & 1.11\\
\hspace{1em}SS-IW & 0.96 & 0.81$^{*}$ & 0.94$^{*}$ & 0.95 & \textbf{0.98} & \textbf{1.03} & \textbf{1.03} & \textbf{0.99} & \textbf{0.99}\\
\hspace{1em}SSNG-IW & 0.98 & 0.83$^{*}$ & 0.95$^{*}$ & 0.97 & 1.01 & 1.08 & 1.09 & 1.07 & 1.06\\
\hspace{1em}Minn-CSV & 0.96 & 0.84$^{*}$ & 0.96 & 1.01 & 1.09 & 1.18 & 1.24 & 1.23 & 1.21\\
\hspace{1em}SS-CSV & \textbf{0.94} & \textbf{0.81}$^{*}$ & \textbf{0.92}$^{*}$ & \textbf{0.95} & 1.01 & 1.06 & 1.10 & 1.07 & 1.07\\
\hspace{1em}SSNG-CSV & 0.95 & 0.82$^{*}$ & 0.94 & 0.98 & 1.05 & 1.14 & 1.19 & 1.17 & 1.16\\
\bottomrule
\end{tabular}
\begin{tablenotes}
\item \textit{Note: } 
\item The forecast horizon $h$ denotes quarters. Negative LPDS entries indicate that the mixed-frequency model is superior in terms of density forecasting and values of the RMSE below 1 indicate better point forecasts. Bold entries show the minimum per column. The benchmark model is a VAR(4) including all 13 variables aggregated to the quarterly frequency using the steady-state prior with a constant error covariance matrix. Two stars ($^{**}$) indicate that the Diebold-Mariano test of equal predictive ability is significant at the 1 percent level, whereas a single star indicates significance at the 10 percent level. The test employs the modifications proposed by \cite{Harvey1997}.
\end{tablenotes}
\end{threeparttable}
\end{table}

\begin{table}[t]
\centering
\fontsize{8}{10}\selectfont
\begin{threeparttable}

\caption{\label{tab:cpiaucsl2}Inflation: Forecast Evaluation (Second Vintage)}
\begin{tabular}{llllllllll}
\toprule
Model & $h = 0$ & $h = 1$ & $h = 2$ & $h = 3$ & $h = 4$ & $h = 5$ & $h = 6$ & $h = 7$ & $h = 8$\\
\midrule
\addlinespace[0.3em]
\multicolumn{10}{l}{\textbf{Relative LPDS} (model in 1st column $-$ benchmark)}\\
\hspace{1em}Minn-IW & 0.06 & -0.07 & -0.02 & -0.01 & 0.01 & -0.02 & -0.02 & -0.02 & -0.03\\
\hspace{1em}SS-IW & 0.06 & -0.09 & -0.00 & -0.00 & 0.03 & -0.00 & -0.01 & -0.00 & -0.01\\
\hspace{1em}SSNG-IW & 0.05 & -0.08 & -0.01 & -0.01 & 0.03 & -0.01 & -0.04 & -0.02$^{*}$ & -0.04\\
\hspace{1em}Minn-CSV & -0.03 & -0.32 & -0.27 & \textbf{-0.31} & -0.32 & -0.31 & -0.31 & -0.36 & -0.34\\
\hspace{1em}SS-CSV & \textbf{-0.05} & \textbf{-0.34} & \textbf{-0.28} & -0.29 & \textbf{-0.33} & \textbf{-0.33} & -0.34 & \textbf{-0.40} & -0.37\\
\hspace{1em}SSNG-CSV & -0.04 & -0.34 & -0.27 & -0.30 & -0.32 & -0.30 & \textbf{-0.35} & -0.40 & \textbf{-0.38}\\
\addlinespace[0.3em]
\multicolumn{10}{l}{\textbf{Relative RMSE} (model in 1st column $/$ benchmark)}\\
\hspace{1em}Minn-IW & 1.02 & 0.98 & 0.99 & 0.99 & 1.00 & 1.00 & 1.00 & 1.00 & 1.00\\
\hspace{1em}SS-IW & \textbf{1.01} & \textbf{0.98} & 0.99 & 1.00 & 1.01 & 1.01 & 1.00 & 1.00 & 1.00\\
\hspace{1em}SSNG-IW & 1.02 & 0.98 & 0.99 & 0.99 & 1.00 & 1.00 & 1.00 & 1.00 & 1.00\\
\hspace{1em}Minn-CSV & 1.04 & 0.99 & 0.98 & 0.97 & \textbf{0.97} & \textbf{0.97} & 0.98$^{*}$ & 0.98$^{*}$ & 0.98\\
\hspace{1em}SS-CSV & 1.04 & 0.99 & \textbf{0.98} & 0.97 & 0.97$^{*}$ & 0.98 & 0.98$^{*}$ & \textbf{0.98}$^{*}$ & \textbf{0.98}\\
\hspace{1em}SSNG-CSV & 1.04 & 0.99 & 0.98 & \textbf{0.97} & 0.97 & 0.97 & \textbf{0.98}$^{*}$ & 0.98$^{*}$ & 0.98\\
\bottomrule
\end{tabular}
\begin{tablenotes}
\item \textit{Note: } 
\item The forecast horizon $h$ denotes months. Negative LPDS entries indicate that the mixed-frequency model is superior in terms of density forecasting and values of the RMSE below 1 indicate better point forecasts. Bold entries show the minimum per column. The benchmark model is a VAR(12) including the ten monthly variables using the steady-state prior with a constant error covariance matrix. Two stars ($^{**}$) indicate that the Diebold-Mariano test of equal predictive ability is significant at the 1 percent level, whereas a single star indicates significance at the 10 percent level. The test employs the modifications proposed by \cite{Harvey1997}.
\end{tablenotes}
\end{threeparttable}
\end{table}

\begin{table}[t]
\centering
\fontsize{8}{10}\selectfont
\begin{threeparttable}

\caption{\label{tab:unrate2}Unemployment: Forecast Evaluation (Second Vintage)}
\begin{tabular}{lllllllll}
\toprule
Model & $h = 1$ & $h = 2$ & $h = 3$ & $h = 4$ & $h = 5$ & $h = 6$ & $h = 7$ & $h = 8$\\
\midrule
\addlinespace[0.3em]
\multicolumn{9}{l}{\textbf{Relative LPDS} (model in 1st column $-$ benchmark)}\\
\hspace{1em}Minn-IW & -0.57 & -0.43 & -0.28 & -0.29 & -0.35$^{*}$ & -0.32 & -0.32$^{*}$ & -0.29\\
\hspace{1em}SS-IW & \textbf{-0.60} & -0.48 & -0.35 & -0.36 & -0.43$^{*}$ & -0.40 & -0.41 & -0.38\\
\hspace{1em}SSNG-IW & -0.56 & -0.41 & -0.25 & -0.25 & -0.30 & -0.26 & -0.26 & -0.21\\
\hspace{1em}Minn-CSV & -0.54 & -0.48 & -0.46 & -0.54 & -0.68 & -0.72 & -0.78 & -0.80\\
\hspace{1em}SS-CSV & -0.56 & \textbf{-0.50} & \textbf{-0.48} & \textbf{-0.56} & \textbf{-0.71} & \textbf{-0.75} & \textbf{-0.82} & \textbf{-0.84}\\
\hspace{1em}SSNG-CSV & -0.53 & -0.46 & -0.44 & -0.51 & -0.65 & -0.69 & -0.76 & -0.78\\
\addlinespace[0.3em]
\multicolumn{9}{l}{\textbf{Relative RMSE} (model in 1st column $/$ benchmark)}\\
\hspace{1em}Minn-IW & 0.74$^{**}$ & 0.81$^{**}$ & 0.86$^{*}$ & 0.87 & 0.86 & 0.87 & 0.88 & 0.90\\
\hspace{1em}SS-IW & \textbf{0.73}$^{**}$ & \textbf{0.79}$^{*}$ & \textbf{0.83}$^{*}$ & \textbf{0.84} & \textbf{0.84} & \textbf{0.85} & \textbf{0.86} & \textbf{0.88}\\
\hspace{1em}SSNG-IW & 0.75$^{**}$ & 0.81$^{**}$ & 0.87$^{*}$ & 0.87 & 0.87 & 0.88 & 0.89 & 0.91\\
\hspace{1em}Minn-CSV & 0.79$^{**}$ & 0.85$^{*}$ & 0.90$^{*}$ & 0.90 & 0.89 & 0.89 & 0.90 & 0.91\\
\hspace{1em}SS-CSV & 0.78$^{**}$ & 0.84$^{*}$ & 0.89$^{*}$ & 0.89 & 0.88 & 0.88 & 0.89 & 0.90\\
\hspace{1em}SSNG-CSV & 0.79$^{**}$ & 0.86$^{*}$ & 0.91 & 0.92 & 0.90 & 0.91 & 0.91 & 0.92\\
\bottomrule
\end{tabular}
\begin{tablenotes}
\item \textit{Note: } 
\item The forecast horizon $h$ denotes months. Negative LPDS entries indicate that the mixed-frequency model is superior in terms of density forecasting and values of the RMSE below 1 indicate better point forecasts. Bold entries show the minimum per column. The benchmark model is a VAR(12) including the ten monthly variables using the steady-state prior with a constant error covariance matrix. Two stars ($^{**}$) indicate that the Diebold-Mariano test of equal predictive ability is significant at the 1 percent level, whereas a single star indicates significance at the 10 percent level. The test employs the modifications proposed by \cite{Harvey1997}.
\end{tablenotes}
\end{threeparttable}
\end{table}

\clearpage
\section{Additional Results}
\label{sec:addres}
This Appendix presents forecast evaluation tables (Table \ref{tab:ceu0500000034}--\ref{tab:tcu}) for the variables not discussed in the main text.

\begin{table}[t]

\centering
\fontsize{8}{10}\selectfont
\begin{threeparttable}
\caption{\label{tab:ceu0500000034}Hours: Forecast Evaluation}
\begin{tabular}{lllllllll}
\toprule
Model & $h = 1$ & $h = 2$ & $h = 3$ & $h = 4$ & $h = 5$ & $h = 6$ & $h = 7$ & $h = 8$\\
\midrule
\addlinespace[0.3em]
\multicolumn{9}{l}{\textbf{Relative LPDS} (model in 1st column $-$ benchmark)}\\
\hspace{1em}Minn-IW & \textbf{-0.11} & -0.12 & \textbf{-0.05} & -0.10 & -0.04$^{*}$ & -0.11 & -0.06 & -0.06\\
\hspace{1em}SS-IW & -0.08 & -0.11 & -0.03 & -0.10 & -0.04$^{*}$ & -0.11 & -0.06 & \textbf{-0.06}\\
\hspace{1em}SSNG-IW & -0.11 & \textbf{-0.12} & -0.04 & -0.12$^{*}$ & -0.04 & -0.11 & \textbf{-0.07} & -0.05\\
\hspace{1em}Minn-CSV & 0.08 & -0.10 & -0.01 & -0.17 & -0.12 & -0.08 & 0.01 & 0.04\\
\hspace{1em}SS-CSV & 0.13 & -0.08 & -0.01 & -0.16 & -0.14 & \textbf{-0.13} & -0.04 & 0.01\\
\hspace{1em}SSNG-CSV & 0.08 & -0.09 & -0.01 & \textbf{-0.19} & \textbf{-0.15} & -0.12 & -0.02 & 0.02\\
\addlinespace[0.3em]
\multicolumn{9}{l}{\textbf{Relative RMSE} (model in 1st column $/$ benchmark)}\\
\hspace{1em}Minn-IW & 0.96$^{*}$ & 0.95$^{*}$ & 0.98 & 0.96$^{*}$ & 0.99 & 0.98 & 0.98 & 0.98\\
\hspace{1em}SS-IW & 0.96$^{*}$ & 0.96$^{*}$ & 0.98 & 0.96$^{*}$ & 0.99$^{*}$ & 0.98 & 0.98 & 0.98\\
\hspace{1em}SSNG-IW & \textbf{0.95}$^{*}$ & \textbf{0.95}$^{*}$ & 0.98 & 0.96$^{*}$ & 0.99$^{*}$ & \textbf{0.98} & \textbf{0.98} & \textbf{0.98}\\
\hspace{1em}Minn-CSV & 0.97 & 0.96 & 0.97 & 0.95$^{*}$ & 0.97 & 0.99 & 0.99 & 0.99\\
\hspace{1em}SS-CSV & 0.98 & 0.97 & 0.98 & 0.95$^{*}$ & 0.98$^{*}$ & 0.98 & 0.98 & 0.99\\
\hspace{1em}SSNG-CSV & 0.97 & 0.96 & \textbf{0.97} & \textbf{0.95}$^{*}$ & \textbf{0.97} & 0.98 & 0.98 & 0.99\\
\bottomrule
\end{tabular}
\begin{tablenotes}
\item \textit{Note: } 
\item The forecast horizon $h$ denotes months. Negative LPDS entries indicate that the mixed-frequency model is superior in terms of density forecasting and values of the RMSE below 1 indicate better point forecasts. Bold entries show the minimum per column. The benchmark model is a VAR(12) including the ten monthly variables using the steady-state prior with a constant error covariance matrix. Two stars ($^{**}$) indicate that the Diebold-Mariano test of equal predictive ability is significant at the 1 percent level, whereas a single star indicates significance at the 10 percent level. The test employs the modifications proposed by \cite{Harvey1997}.
\end{tablenotes}
\end{threeparttable}
\end{table}

\begin{table}[t]

\centering
\fontsize{8}{10}\selectfont
\begin{threeparttable}
\caption{\label{tab:indpro}Industrial Production: Forecast Evaluation}
\begin{tabular}{llllllllll}
\toprule
Model & $h = 0$ & $h = 1$ & $h = 2$ & $h = 3$ & $h = 4$ & $h = 5$ & $h = 6$ & $h = 7$ & $h = 8$\\
\midrule
\addlinespace[0.3em]
\multicolumn{10}{l}{\textbf{Relative LPDS} (model in 1st column $-$ benchmark)}\\
\hspace{1em}Minn-IW & -0.06 & -0.04 & -0.16$^{*}$ & -0.16$^{*}$ & -0.07$^{*}$ & -0.06 & -0.12 & -0.04 & -0.08\\
\hspace{1em}SS-IW & -0.05 & -0.02 & -0.15$^{*}$ & -0.16$^{*}$ & -0.08$^{*}$ & -0.06 & -0.12 & -0.06 & -0.10\\
\hspace{1em}SSNG-IW & -0.03 & -0.03 & -0.15$^{*}$ & -0.15$^{*}$ & -0.06$^{*}$ & -0.06 & -0.10 & -0.06 & -0.06\\
\hspace{1em}Minn-CSV & -0.36 & \textbf{-0.27} & -0.36 & -0.35$^{*}$ & -0.32 & -0.28 & -0.30 & -0.23 & -0.23\\
\hspace{1em}SS-CSV & -0.37 & -0.25 & -0.35 & -0.34$^{*}$ & -0.33 & -0.30 & \textbf{-0.33} & \textbf{-0.27} & \textbf{-0.27}\\
\hspace{1em}SSNG-CSV & \textbf{-0.38}$^{*}$ & -0.27 & \textbf{-0.36} & \textbf{-0.35}$^{*}$ & \textbf{-0.34} & \textbf{-0.31} & -0.33 & -0.27 & -0.26\\
\addlinespace[0.3em]
\multicolumn{10}{l}{\textbf{Relative RMSE} (model in 1st column $/$ benchmark)}\\
\hspace{1em}Minn-IW & 0.92$^{*}$ & 0.98 & 0.95$^{*}$ & 0.95$^{*}$ & 0.98$^{*}$ & 0.99 & \textbf{0.97} & 0.99 & 0.99\\
\hspace{1em}SS-IW & 0.91$^{*}$ & 0.98 & 0.95$^{*}$ & 0.95$^{*}$ & 0.98 & 0.99 & 0.97 & 0.99 & \textbf{0.98}\\
\hspace{1em}SSNG-IW & 0.92$^{*}$ & 0.98 & 0.96$^{*}$ & 0.95$^{*}$ & 0.98$^{*}$ & 0.99 & 0.97 & \textbf{0.99} & 0.99\\
\hspace{1em}Minn-CSV & 0.91$^{**}$ & \textbf{0.97} & \textbf{0.94}$^{*}$ & 0.94$^{*}$ & 0.98$^{*}$ & 0.98$^{*}$ & 0.98 & 1.01 & 1.01\\
\hspace{1em}SS-CSV & 0.91$^{**}$ & 0.97 & 0.95$^{*}$ & 0.94$^{*}$ & 0.98 & 0.99$^{*}$ & 0.98 & 1.00 & 1.00\\
\hspace{1em}SSNG-CSV & \textbf{0.91}$^{**}$ & 0.97 & 0.94$^{*}$ & \textbf{0.94}$^{*}$ & \textbf{0.98}$^{*}$ & \textbf{0.98}$^{*}$ & 0.98 & 1.00 & 1.00$^{*}$\\
\bottomrule
\end{tabular}
\begin{tablenotes}
\item \textit{Note: } 
\item The forecast horizon $h$ denotes months. Negative LPDS entries indicate that the mixed-frequency model is superior in terms of density forecasting and values of the RMSE below 1 indicate better point forecasts. Bold entries show the minimum per column. The benchmark model is a VAR(12) including the ten monthly variables using the steady-state prior with a constant error covariance matrix. Two stars ($^{**}$) indicate that the Diebold-Mariano test of equal predictive ability is significant at the 1 percent level, whereas a single star indicates significance at the 10 percent level. The test employs the modifications proposed by \cite{Harvey1997}.
\end{tablenotes}
\end{threeparttable}
\end{table}

\begin{table}[t]

\centering
\fontsize{8}{10}\selectfont
\begin{threeparttable}
\caption{\label{tab:payems}Non-Farm Employment: Forecast Evaluation}
\begin{tabular}{lllllllll}
\toprule
Model & $h = 1$ & $h = 2$ & $h = 3$ & $h = 4$ & $h = 5$ & $h = 6$ & $h = 7$ & $h = 8$\\
\midrule
\addlinespace[0.3em]
\multicolumn{9}{l}{\textbf{Relative LPDS} (model in 1st column $-$ benchmark)}\\
\hspace{1em}Minn-IW & -0.13$^{**}$ & -0.24$^{**}$ & -0.21$^{*}$ & -0.18$^{*}$ & -0.18 & -0.18 & -0.15 & -0.18\\
\hspace{1em}SS-IW & -0.12$^{**}$ & -0.24$^{**}$ & -0.22$^{*}$ & -0.20$^{*}$ & -0.20 & -0.21 & -0.19 & -0.23\\
\hspace{1em}SSNG-IW & -0.14$^{**}$ & -0.24$^{**}$ & -0.21$^{*}$ & -0.19$^{*}$ & -0.18 & -0.20 & -0.15 & -0.18\\
\hspace{1em}Minn-CSV & -0.39$^{**}$ & -0.41$^{**}$ & -0.40$^{*}$ & \textbf{-0.36}$^{*}$ & \textbf{-0.31} & -0.31 & \textbf{-0.27} & -0.29\\
\hspace{1em}SS-CSV & -0.38$^{**}$ & -0.41$^{**}$ & -0.40$^{*}$ & -0.35$^{*}$ & -0.31 & \textbf{-0.31} & -0.27 & \textbf{-0.30}\\
\hspace{1em}SSNG-CSV & \textbf{-0.39}$^{**}$ & \textbf{-0.42}$^{**}$ & \textbf{-0.40}$^{*}$ & -0.35$^{*}$ & -0.31 & -0.31 & -0.26 & -0.29\\
\addlinespace[0.3em]
\multicolumn{9}{l}{\textbf{Relative RMSE} (model in 1st column $/$ benchmark)}\\
\hspace{1em}Minn-IW & 0.93$^{*}$ & 0.88$^{*}$ & 0.91$^{*}$ & 0.92 & 0.93 & 0.93 & 0.95 & 0.94\\
\hspace{1em}SS-IW & 0.93$^{*}$ & 0.88$^{*}$ & 0.90$^{*}$ & 0.91$^{*}$ & 0.92 & 0.92 & 0.94 & 0.92\\
\hspace{1em}SSNG-IW & 0.93$^{*}$ & 0.88$^{*}$ & 0.90$^{*}$ & 0.92$^{*}$ & 0.93 & 0.93 & 0.95 & 0.94\\
\hspace{1em}Minn-CSV & 0.89$^{**}$ & \textbf{0.87}$^{*}$ & \textbf{0.88}$^{*}$ & \textbf{0.90}$^{*}$ & \textbf{0.91} & \textbf{0.91} & 0.93 & 0.93\\
\hspace{1em}SS-CSV & 0.90$^{**}$ & 0.87$^{*}$ & 0.88$^{*}$ & 0.90$^{*}$ & 0.91 & 0.91 & \textbf{0.93} & \textbf{0.92}\\
\hspace{1em}SSNG-CSV & \textbf{0.89}$^{**}$ & 0.87$^{*}$ & 0.88$^{*}$ & 0.90$^{*}$ & 0.91 & 0.91 & 0.93 & 0.93\\
\bottomrule
\end{tabular}
\begin{tablenotes}
\item \textit{Note: } 
\item The forecast horizon $h$ denotes months. Negative LPDS entries indicate that the mixed-frequency model is superior in terms of density forecasting and values of the RMSE below 1 indicate better point forecasts. Bold entries show the minimum per column. The benchmark model is a VAR(12) including the ten monthly variables using the steady-state prior with a constant error covariance matrix. Two stars ($^{**}$) indicate that the Diebold-Mariano test of equal predictive ability is significant at the 1 percent level, whereas a single star indicates significance at the 10 percent level. The test employs the modifications proposed by \cite{Harvey1997}.
\end{tablenotes}
\end{threeparttable}
\end{table}

\begin{table}[t]

\centering
\fontsize{8}{10}\selectfont
\begin{threeparttable}
\caption{\label{tab:pce}Consumption: Forecast Evaluation}
\begin{tabular}{llllllllll}
\toprule
Model & $h = 0$ & $h = 1$ & $h = 2$ & $h = 3$ & $h = 4$ & $h = 5$ & $h = 6$ & $h = 7$ & $h = 8$\\
\midrule
\addlinespace[0.3em]
\multicolumn{10}{l}{\textbf{Relative LPDS} (model in 1st column $-$ benchmark)}\\
\hspace{1em}Minn-IW & -0.05$^{*}$ & -0.03 & -0.00 & 0.00 & 0.00 & \textbf{-0.00} & 0.00 & 0.01 & 0.00\\
\hspace{1em}SS-IW & -0.07$^{*}$ & -0.04 & -0.00 & 0.01 & 0.01 & 0.00 & 0.00 & 0.01 & -0.00\\
\hspace{1em}SSNG-IW & -0.06$^{*}$ & -0.04 & -0.01 & \textbf{-0.00} & \textbf{0.00} & 0.00 & \textbf{0.00} & \textbf{0.00} & \textbf{-0.00}\\
\hspace{1em}Minn-CSV & -0.23$^{*}$ & -0.15$^{*}$ & -0.04 & 0.03 & 0.08 & 0.12 & 0.15 & 0.18 & 0.19\\
\hspace{1em}SS-CSV & \textbf{-0.23}$^{*}$ & \textbf{-0.15}$^{*}$ & \textbf{-0.05} & 0.04 & 0.09 & 0.13 & 0.16 & 0.19 & 0.20\\
\hspace{1em}SSNG-CSV & -0.22$^{*}$ & -0.14$^{*}$ & -0.04 & 0.03 & 0.09 & 0.13 & 0.17 & 0.19 & 0.21\\
\addlinespace[0.3em]
\multicolumn{10}{l}{\textbf{Relative RMSE} (model in 1st column $/$ benchmark)}\\
\hspace{1em}Minn-IW & 0.98$^{*}$ & 0.98 & 1.01 & 1.01 & 1.00 & 1.01 & 1.01 & 1.01 & 1.00\\
\hspace{1em}SS-IW & \textbf{0.96}$^{*}$ & 0.97 & 1.01 & 1.01 & 1.01 & 1.01 & 1.00 & 1.00 & 0.99\\
\hspace{1em}SSNG-IW & 0.98$^{*}$ & 0.98 & \textbf{1.00} & 1.00 & 1.00 & 1.01 & 1.00 & 1.00 & 0.99$^{*}$\\
\hspace{1em}Minn-CSV & 1.01 & 0.97 & 1.03 & 0.99 & 1.00 & 1.00 & 1.00 & 1.02 & 1.01\\
\hspace{1em}SS-CSV & 0.98 & \textbf{0.96} & 1.02 & \textbf{0.99} & \textbf{0.99} & \textbf{0.99} & \textbf{0.98} & \textbf{0.99} & \textbf{0.99}\\
\hspace{1em}SSNG-CSV & 1.01 & 0.97 & 1.02 & 0.99 & 1.00 & 0.99 & 1.00 & 1.01 & 1.00\\
\bottomrule
\end{tabular}
\begin{tablenotes}
\item \textit{Note: } 
\item The forecast horizon $h$ denotes months. Negative LPDS entries indicate that the mixed-frequency model is superior in terms of density forecasting and values of the RMSE below 1 indicate better point forecasts. Bold entries show the minimum per column. The benchmark model is a VAR(12) including the ten monthly variables using the steady-state prior with a constant error covariance matrix. Two stars ($^{**}$) indicate that the Diebold-Mariano test of equal predictive ability is significant at the 1 percent level, whereas a single star indicates significance at the 10 percent level. The test employs the modifications proposed by \cite{Harvey1997}.
\end{tablenotes}
\end{threeparttable}
\end{table}

\begin{table}[t]

\centering
\fontsize{8}{10}\selectfont
\begin{threeparttable}
\caption{\label{tab:sp500}S\&P500: Forecast Evaluation}
\begin{tabular}{lllllllll}
\toprule
Model & $h = 1$ & $h = 2$ & $h = 3$ & $h = 4$ & $h = 5$ & $h = 6$ & $h = 7$ & $h = 8$\\
\midrule
\addlinespace[0.3em]
\multicolumn{9}{l}{\textbf{Relative LPDS} (model in 1st column $-$ benchmark)}\\
\hspace{1em}Minn-IW & -0.10 & -0.06$^{*}$ & -0.03 & -0.06 & -0.07 & \textbf{-0.04} & \textbf{-0.05} & \textbf{-0.05}\\
\hspace{1em}SS-IW & -0.09 & -0.01 & 0.03 & -0.01 & -0.01 & 0.02 & 0.01 & 0.01\\
\hspace{1em}SSNG-IW & -0.10 & -0.04$^{*}$ & -0.01 & -0.05 & -0.06 & -0.02 & -0.03$^{*}$ & -0.02\\
\hspace{1em}Minn-CSV & -0.20 & \textbf{-0.16} & \textbf{-0.11} & \textbf{-0.08} & -0.07 & -0.04 & -0.05 & -0.03\\
\hspace{1em}SS-CSV & -0.19 & -0.15 & -0.08 & -0.06 & -0.06 & -0.02 & -0.02 & -0.00\\
\hspace{1em}SSNG-CSV & \textbf{-0.21} & -0.16 & -0.11 & -0.07 & \textbf{-0.08} & -0.04 & -0.03 & -0.01\\
\addlinespace[0.3em]
\multicolumn{9}{l}{\textbf{Relative RMSE} (model in 1st column $/$ benchmark)}\\
\hspace{1em}Minn-IW & \textbf{0.94} & \textbf{0.98} & \textbf{0.98} & \textbf{0.97} & \textbf{0.96} & \textbf{0.98} & \textbf{0.98} & \textbf{0.98}\\
\hspace{1em}SS-IW & 0.95 & 1.00 & 1.02 & 1.00 & 1.00 & 1.02 & 1.01 & 1.01\\
\hspace{1em}SSNG-IW & 0.94 & 0.99 & 0.99 & 0.98 & 0.97 & 0.99 & 0.99 & 0.99\\
\hspace{1em}Minn-CSV & 0.96 & 0.98 & 0.99 & 0.97 & 0.97 & 0.99 & 0.98 & 0.98\\
\hspace{1em}SS-CSV & 0.97 & 0.99 & 1.01 & 0.99 & 0.99 & 1.00 & 1.00 & 0.99\\
\hspace{1em}SSNG-CSV & 0.96 & 0.98 & 0.99 & 0.98 & 0.97 & 0.99 & 0.99 & 0.99\\
\bottomrule
\end{tabular}
\begin{tablenotes}
\item \textit{Note: } 
\item The forecast horizon $h$ denotes months. Negative LPDS entries indicate that the mixed-frequency model is superior in terms of density forecasting and values of the RMSE below 1 indicate better point forecasts. Bold entries show the minimum per column. The benchmark model is a VAR(12) including the ten monthly variables using the steady-state prior with a constant error covariance matrix. Two stars ($^{**}$) indicate that the Diebold-Mariano test of equal predictive ability is significant at the 1 percent level, whereas a single star indicates significance at the 10 percent level. The test employs the modifications proposed by \cite{Harvey1997}.
\end{tablenotes}
\end{threeparttable}
\end{table}

\begin{table}[t]
\centering
\fontsize{8}{10}\selectfont
\begin{threeparttable}

\caption{\label{tab:t10yff}Bond Spread: Forecast Evaluation}
\begin{tabular}{lllllllll}
\toprule
Model & $h = 1$ & $h = 2$ & $h = 3$ & $h = 4$ & $h = 5$ & $h = 6$ & $h = 7$ & $h = 8$\\
\midrule
\addlinespace[0.3em]
\multicolumn{9}{l}{\textbf{Relative LPDS} (model in 1st column $-$ benchmark)}\\
\hspace{1em}Minn-IW & -0.82$^{**}$ & -0.36$^{**}$ & -0.18$^{**}$ & -0.10 & -0.08 & -0.08 & -0.08 & -0.08\\
\hspace{1em}SS-IW & -0.83$^{**}$ & -0.37$^{**}$ & -0.19$^{**}$ & -0.11$^{*}$ & -0.08 & -0.08 & -0.06 & -0.05\\
\hspace{1em}SSNG-IW & -0.82$^{**}$ & -0.37$^{**}$ & -0.19$^{**}$ & -0.12$^{*}$ & \textbf{-0.09} & \textbf{-0.10} & \textbf{-0.09} & \textbf{-0.09}\\
\hspace{1em}Minn-CSV & \textbf{-1.06}$^{**}$ & \textbf{-0.58}$^{**}$ & \textbf{-0.33}$^{*}$ & \textbf{-0.17} & -0.05 & 0.01 & 0.06 & 0.08\\
\hspace{1em}SS-CSV & -1.05$^{**}$ & -0.58$^{**}$ & -0.33$^{*}$ & -0.16 & -0.04 & 0.03 & 0.08 & 0.10\\
\hspace{1em}SSNG-CSV & -1.05$^{**}$ & -0.57$^{*}$ & -0.33$^{*}$ & -0.16 & -0.04 & 0.02 & 0.07 & 0.09\\
\addlinespace[0.3em]
\multicolumn{9}{l}{\textbf{Relative RMSE} (model in 1st column $/$ benchmark)}\\
\hspace{1em}Minn-IW & 0.65$^{**}$ & 0.84$^{*}$ & 0.94$^{*}$ & 0.98 & 0.99 & 0.99 & 0.98 & 0.98\\
\hspace{1em}SS-IW & 0.64$^{**}$ & 0.83$^{**}$ & 0.92$^{*}$ & \textbf{0.96} & 0.97 & 0.98 & 0.99 & 0.99\\
\hspace{1em}SSNG-IW & 0.65$^{**}$ & 0.83$^{**}$ & 0.92$^{*}$ & 0.96 & \textbf{0.97} & \textbf{0.97} & \textbf{0.97} & \textbf{0.97}\\
\hspace{1em}Minn-CSV & 0.63$^{*}$ & 0.82$^{*}$ & 0.92 & 1.00 & 1.04 & 1.06 & 1.08 & 1.09\\
\hspace{1em}SS-CSV & \textbf{0.62}$^{**}$ & \textbf{0.81}$^{*}$ & \textbf{0.91} & 0.98 & 1.03 & 1.06 & 1.08 & 1.09\\
\hspace{1em}SSNG-CSV & 0.63$^{*}$ & 0.82$^{*}$ & 0.92 & 0.99 & 1.04 & 1.06 & 1.08 & 1.09\\
\bottomrule
\end{tabular}
\begin{tablenotes}
\item \textit{Note: } 
\item The forecast horizon $h$ denotes months. Negative LPDS entries indicate that the mixed-frequency model is superior in terms of density forecasting and values of the RMSE below 1 indicate better point forecasts. Bold entries show the minimum per column. The benchmark model is a VAR(12) including the ten monthly variables using the steady-state prior with a constant error covariance matrix. Two stars ($^{**}$) indicate that the Diebold-Mariano test of equal predictive ability is significant at the 1 percent level, whereas a single star indicates significance at the 10 percent level. The test employs the modifications proposed by \cite{Harvey1997}.
\end{tablenotes}
\end{threeparttable}
\end{table}

\begin{table}[t]

\centering
\fontsize{8}{10}\selectfont
\begin{threeparttable}
\caption{\label{tab:tcu}Capacity Utilization: Forecast Evaluation}
\begin{tabular}{llllllllll}
\toprule
Model & $h = 0$ & $h = 1$ & $h = 2$ & $h = 3$ & $h = 4$ & $h = 5$ & $h = 6$ & $h = 7$ & $h = 8$\\
\midrule
\addlinespace[0.3em]
\multicolumn{10}{l}{\textbf{Relative LPDS} (model in 1st column $-$ benchmark)}\\
\hspace{1em}Minn-IW & 1.40 & 0.39 & 0.07 & -0.11 & -0.16 & -0.23 & -0.28 & -0.30 & -0.34\\
\hspace{1em}SS-IW & \textbf{1.30} & \textbf{0.37} & 0.09 & -0.10 & -0.19 & -0.27 & -0.34 & -0.38 & -0.44\\
\hspace{1em}SSNG-IW & 1.37 & 0.42 & 0.15 & -0.03 & -0.09 & -0.17 & -0.24 & -0.26 & -0.30\\
\hspace{1em}Minn-CSV & 3.53 & 0.77 & -0.08 & -0.43 & -0.62 & -0.76 & -0.88 & -0.93 & -1.00\\
\hspace{1em}SS-CSV & 3.39 & 0.70 & -0.12 & -0.46 & -0.68 & -0.84 & \textbf{-0.98} & \textbf{-1.06} & \textbf{-1.14}\\
\hspace{1em}SSNG-CSV & 3.48 & 0.74 & \textbf{-0.12} & \textbf{-0.47} & \textbf{-0.68} & \textbf{-0.84} & -0.98 & -1.05 & -1.13\\
\addlinespace[0.3em]
\multicolumn{10}{l}{\textbf{Relative RMSE} (model in 1st column $/$ benchmark)}\\
\hspace{1em}Minn-IW & 0.96$^{**}$ & 0.95$^{*}$ & 0.94$^{*}$ & 0.93$^{*}$ & 0.93$^{*}$ & 0.94 & 0.94 & 0.95 & 0.95\\
\hspace{1em}SS-IW & \textbf{0.95}$^{**}$ & \textbf{0.95}$^{*}$ & \textbf{0.93}$^{*}$ & \textbf{0.92}$^{*}$ & \textbf{0.92}$^{*}$ & \textbf{0.92} & \textbf{0.92} & \textbf{0.92} & \textbf{0.93}\\
\hspace{1em}SSNG-IW & 0.95$^{**}$ & 0.95$^{*}$ & 0.94$^{*}$ & 0.94$^{*}$ & 0.94$^{*}$ & 0.94 & 0.94 & 0.95 & 0.95\\
\hspace{1em}Minn-CSV & 0.98$^{*}$ & 0.98 & 0.97 & 0.97 & 0.98 & 0.99$^{*}$ & 0.99$^{*}$ & 1.00 & 1.01\\
\hspace{1em}SS-CSV & 0.97$^{**}$ & 0.97$^{*}$ & 0.96$^{*}$ & 0.96 & 0.97 & 0.97 & 0.97 & 0.98 & 0.98$^{*}$\\
\hspace{1em}SSNG-CSV & 0.97$^{*}$ & 0.97$^{*}$ & 0.96$^{*}$ & 0.96 & 0.97 & 0.97 & 0.97 & 0.98 & 0.99$^{*}$\\
\bottomrule
\end{tabular}
\begin{tablenotes}
\item \textit{Note: } 
\item The forecast horizon $h$ denotes months. Negative LPDS entries indicate that the mixed-frequency model is superior in terms of density forecasting and values of the RMSE below 1 indicate better point forecasts. Bold entries show the minimum per column. The benchmark model is a VAR(12) including the ten monthly variables using the steady-state prior with a constant error covariance matrix. Two stars ($^{**}$) indicate that the Diebold-Mariano test of equal predictive ability is significant at the 1 percent level, whereas a single star indicates significance at the 10 percent level. The test employs the modifications proposed by \cite{Harvey1997}.
\end{tablenotes}
\end{threeparttable}
\end{table}

\end{document}